\documentclass[10pt,english,british,tightenlines,eqsecnum,
floats,aps,amsmath,amssymb,nofootinbib,superscriptaddress,prd,showpacs,showkeys]
              {revtex4}
\usepackage[latin9]{inputenc}
\usepackage{color}
\usepackage{babel}
\usepackage{amsmath}
\usepackage{amssymb}
\usepackage{graphicx}
\usepackage{esint}
\usepackage[unicode=true,pdfusetitle,
 bookmarks=true,bookmarksnumbered=false,bookmarksopen=false,
 breaklinks=false,pdfborder={0 0 1},backref=false,colorlinks=false]
 {hyperref}

\makeatletter
\@ifundefined{textcolor}{}
{%
 \definecolor{BLACK}{gray}{0}
 \definecolor{WHITE}{gray}{1}
 \definecolor{RED}{rgb}{1,0,0}
 \definecolor{GREEN}{rgb}{0,1,0}
 \definecolor{BLUE}{rgb}{0,0,1}
 \definecolor{CYAN}{cmyk}{1,0,0,0}
 \definecolor{MAGENTA}{cmyk}{0,1,0,0}
 \definecolor{YELLOW}{cmyk}{0,0,1,0}
}

\@ifundefined{date}{}{\date{March 16, 2014}}
\usepackage{babel}

\usepackage{euscript}\usepackage{epsfig}

\usepackage{amsfonts}

\usepackage{fancyhdr}

\makeatother

\begin{document}

\title{Modified Brans--Dicke Theory in Arbitrary Dimensions}

\author{S. M. M. Rasouli}

\email{mrasouli@ubi.pt}

\affiliation{Departamento de F\'{i}sica, Universidade da Beira Interior, Rua Marqu\^{e}s d'Avila
e Bolama, 6200 Covilh\~{a}, Portugal}

\affiliation{Centro de Matem\'{a}tica e Aplica\c{c}\~{o}es (CMA - UBI),
Universidade da Beira Interior, Rua Marqu\^{e}s d'Avila
e Bolama, 6200 Covilh\~{a}, Portugal}
\author{Mehrdad Farhoudi}

\email{m-farhoudi@sbu.ac.ir}

\affiliation{Department of Physics, Shahid Beheshti
             University, G.C., Evin, Tehran 19839, Iran}

\author{Paulo Vargas Moniz}

\email{pmoniz@ubi.pt}

\affiliation{Departamento de F\'{i}sica, Universidade da Beira Interior, Rua Marqu\^{e}s d'Avila
e Bolama, 6200 Covilh\~{a}, Portugal}

\affiliation{Centro de Matem\'{a}tica e Aplica\c{c}\~{o}es (CMA - UBI),
Universidade da Beira Interior, Rua Marqu\^{e}s d'Avila
e Bolama, 6200 Covilh\~{a}, Portugal}

\begin{abstract}
Within an algebraic framework, used
to construct the induced--matter--theory~(IMT) setting,
in $(D+1)$--dimensional
Brans--Dicke~(BD) scenario,
we obtain a modified BD theory (MBDT) in $D$
dimensions. Being more specific, from the $(D+1)$--dimensional field
equations, a $D$--dimensional BD theory, bearing new features, is extracted by means of a
suitable dimensional reduction onto a hypersurface orthogonal to the extra
dimension. In particular, the BD scalar field in such
$D$--dimensional theory has a self--interacting potential, which
can be suitably interpreted as produced by the extra dimension.
Subsequently, as an application to cosmology, we consider
 an extended spatially flat FLRW geometry in a
$(D+1)$--dimensional space--time.
After obtaining the power--law solutions in the bulk,
we proceed to construct the corresponding physics, by means of the
induced MBDT procedure, on the $D$--dimensional hypersurface. We then
 contrast the resulted solutions
 (for different phases of the universe) with those usually extracted from the
 conventional GR and BD theories in view of current ranges for cosmological parameters.
 We show that the induced perfect fluid background and the induced scalar potential
 can be employed, within some limits,
 for describing different epochs of the universe.
 Finally, we comment on the observational viability of such a model.
 \end{abstract}

\medskip



\maketitle

\section{Introduction}
\label{int} \indent

Models of the universe in more than four dimensions have been
widely investigated. Kaluza-Klein~(KK) type
theories~\cite{Kaluza21,Klein26,OW97},
ten--dimensional
%~\cite{Paulo.book1,Paulo.book2}
and
eleven--dimensional supergravity~\cite{FP12.book} as well as string
theories~\cite{DNP86,GSW.book} are well--known examples.
Multidimensional Brane--World models, space--time--matter or IMT
scenarios~\cite{stm99,scalarbook,5Dwesson06,pav2006} in five
dimensions, seeking the unification of matter and geometry,
constitute other settings with additional spatial dimensions.

Unifying electromagnetism with gravity has been investigated by Kaluza by admitting
Einstein general relativity (GR) in a five--dimensional
space--time under three key assumptions~\cite{OW97}: (i) absence of matter in
higher dimensional space--time, (ii) defining the geometrical quantities exactly the same as they were in GR
and (iii) omitting the derivatives with respect to the extra coordinate (cylinder condition).
Since Kaluza's procedure in 1921, compactified,
projective and noncompactified versions have been studied as interesting
approaches to higher dimensional unification, in which, at least
one of the Kaluza's key assumptions has been modified.
%in each of these three mentioned approaches.
Among these mentioned versions, we will consider the third. In fact, the IMT
is one of these developed procedures.
% in the latter approach by accepting its main assumption.
In the IMT, the presence of extra dimensions is also elevated to the
hypothesis that matter in a four--dimensional space--time has a purely
geometric origin. More precisely, it has been
proposed~\cite{stm99,5Dwesson06,PW92,OW97} that one {\it large}
extra dimension is required to obtain a consistent description, at
the macroscopic level, of the properties of matter as observed in
the four--dimensional space--time of GR.
%In the context of cosmology, the IMT has been applied to study the
%properties of such matter, comparing them with
%the results reported by the
The IMT has been employed in the cosmological context,
where concrete scenarios have been appreciated in view of recent
cosmological observational
data~\cite{stm99,DRJ09,RJ10}. The application of the IMT
framework to arbitrary dimensions has been performed
in~\cite{RRT95}, relating a vacuum $(D+1)$--dimensional solution
to a $D$--dimensional GR space--time, with induced matter
sources\rlap.\footnote{Geometrically generated by means of
the dimensional reduction process.}\
 This approach has also been
employed to obtain lower dimensional gravity from a
four--dimensional space--time description.

Another generalization of the IMT, in which the role of GR as a
fundamental underlying theory is replaced by the BD theory of
gravity, has also been investigated~\cite{ARB07,Ponce1,Ponce2}.
The BD theory is an extension of GR, in which the Newton
gravitational constant is substituted, in the Jordan frame, by a non--minimally coupled
scalar field~\cite{BD61,D62}. In this latter application of
the IMT, it has been shown that five--dimensional BD
vacuum\footnote{From now on, we call ``vacuum'' to a situation
where there is~not any other type of ordinary matter, with  the BD
scalar field being the only formal ``source'' of gravity.
We should also note that in~\cite{Ponce2}, the inducing procedure has been started
from a very general BD field equations, rather than the vacuum space--time.}
  equations, when reduced to four
dimensions, induce a modified four--dimensional BD theory. This
feature is of some relevance. In fact, despite of some
(``conventional'') versions of a four--dimensional BD setting,
where a few assumptions have been advocated in order to obtain an
accelerating cosmos\rlap,\footnote{For example, assuming the BD
coupling parameter $\omega$ as a function of the time~\cite{BP01},
or introducing a time--dependent cosmological term~\cite{MC07}
and/or adding a particular kind of scalar potential to
the Lagrangian (or without considering any scalar potential) by
assuming a fluid with dissipative pressure~\cite{SS01,SSS01}.
Further, in~\cite{SS03}, the authors derived the accelerating universe
in the BD theory by assuming a scalar potential
compatible with the power--law expansion of the universe.
Also, in~\cite{BM00}, it has been shown that the BD setting
with a quadratic self--coupling of the BD scalar field and a
negative $\omega$ leads to accelerated expansion solutions.}
 the mentioned IMT setup
within a BD theory~\cite{ARB07,Ponce1,Ponce2} provides a
more appealing perspective based on a fundamental concept. More concretely, in the context of
spatially flat Friedmann--Lema\^{\i}tre--Robertson--Walker (FLRW) cosmology, by
employing the Wesson idea~\cite{stm99,5Dwesson06,PW92,OW97} for
the BD theory, it has then been shown~\cite{Ponce1,Ponce2} that
the subsequent self--interacting scalar potential (geometrically
due to the extra dimension) and the induced matter lead to
 cosmological acceleration of the matter dominated universe.
 Furthermore, the generalized Bianchi
type~I~\cite{RFS11} and FLRW~\cite{BFS11}
models have been studied in this scenario.
Our intention
in this work will be to generalize the simplest scalar--tensor
gravity model, the BD theory, under some
of the main assumptions of the IMT, and then critical
studying a spatially flat FLRW cosmological model in the
extracted gravity model.

In the other reduced BD setting (different from~\cite{Ponce1,Ponce2}), that is also based on a
fundamental concept,
a five--dimensional manifold with a compact and sufficiently small fifth dimension
(cylindiricity condition) has been assumed and then the five--dimensional BD equations reduced
on a hypersurface orthogonal to the extra dimension~\cite{qiang2005,qiang2009}.
%in which the
%fluid therein does not move along the fifth dimension.
By assuming a few constraints especially on the matter content in five-dimensional space--time,
the four--metric in this four--dimensional reduced theory is coupled with two scalar fields,
which are responsible for the accelerated expansion of the universe.

In the context conveyed in the previous paragraphs, the objective
of our work is to generalize the IMT formulation of the BD setting
towards any arbitrary $D$--dimensional
%\footnote{The dimensional
%\bl{reduction} of a five--dimensional BD theory, with matter content,
%on a hypersurface orthogonal to the extra dimension (recovering a
%modified BD theory in four dimensions) has previously been
%performed in~\cite{qiang2005}.}\
 space--time. Our work is
organized as follows. In Section~\ref{Set up}, we derive the BD
field equations in $(D+1)$ dimensions and then, by applying a
dimensional reduction procedure, within an IMT framework, we construct the
MBDT on a hypersurface. Moreover, in this section, the
$(D+1)$--dimensional field equations lead to a very specific
$D$--dimensional\footnote{By $D$--dimensional, we mean
$((D-1),1)$--dimensional space--time.}\
 BD theory, where new dynamical ingredients are present, namely, an effective induced
self--interacting scalar potential. Subsequently, we investigate
cosmological applications. More concretely, in
Section~\ref{OT-solution}, we discuss exact solutions of BD
cosmology in a ($D+1$)--dimensional vacuum space--time. Then, in
Section~\ref{OT-reduced}, by means of the MBDT--IMT framework, we
study the reduced $D$--dimensional cosmological solutions.
We analyze them for different ranges of the equation of state parameter
in a four--dimensional space-time and, subsequently,
compare our results with recent constraints on the BD theory~\cite{LWC13} based on the new
cosmological data (e.g. Planck~\cite{Planck.XVI}) as
well as obtained results in the context of the standard BD theory, e.g.,~\cite{BP01}-\cite{BM00}.
%Besides standard expanding scenarios, we indicate an interesting accelerating case as well as
%decelerated expansion which is described with a radiation dominated universe.
Finally, we present our conclusions in Section~\ref{conclusion}.
In Appendix~\ref{App.A}, we show that a
$(D+1)$--dimensional BD theory
 can be derived from the simplest
 version of a KK theory
 %with a single dilaton field
 in a $(D+1+d)$--dimensional space--time.
 % together with accepting the KK three key assumptions.

\section{$D$--Dimensional Brans--Dicke Theory From $(D+1)$ Dimensions}
\label{Set up}
\indent

The action for the $(D+1)$--dimensional BD theory, in the Jordan
frame, is written  as
\begin{equation}\label{(D+1)-action}
{\cal S}^{^{(D+1)}}=\int d^{^{D+1}}x \sqrt{\Bigl|{}{\cal
G}^{^{(D+1)}}\Bigr|} \,\left[\phi
R^{^{(D+1)}}-\frac{\omega}{\phi}\, {\cal
G}^{ab}\,(\nabla_a\phi)(\nabla_b\phi)+16\pi\,
L\!^{^{(D+1)}}_{_{\rm matt}}\right],
\end{equation}
where $\phi$ is the BD scalar field, $\omega$ is an adjustable
dimensionless parameter called the BD coupling
parameter\rlap,\footnote{Usually, concerning the possibility of
applying a conformal transformation to bring the theory from the
Jordan frame to the Einstein frame, the BD coupling parameter is
assumed to be $\omega>\!-(D-1)/(D-2)$ for a $D$--dimensional
space--time~\cite{Faraoni.book}.}\
 the Latin indices run from zero
to $D$, $R^{^{(D+1)}}$ is the curvature scalar associated with the
$(D+1)$--dimensional space--time metric ${\cal G}_{{ab}}$, ${\cal
G}^{^{(D+1)}}$ is the determinant of the metric and $\nabla_a$
denotes the covariant derivative in $(D+1)$--dimensional
space--time. The Lagrangian $L^{^{(D+1)}}_{_{\rm matt}}$ describes
 ordinary matter in $(D+1)$--dimensional space--time, which
depends on the metric and other matter fields except on $\phi$,
and we have chosen $c=1$.

The variation of action (\ref{(D+1)-action}), with respect to the
metric and the scalar field, gives the equations
\begin{equation}\label{(D+1)-equation-1}
G^{^{(D+1)}}_{ab}=\frac{8\pi}{\phi}\,T^{^{(D+1)}}_{ab}+\frac{\omega}{\phi^{2}}
\left[(\nabla_a\phi)(\nabla_b\phi)-\frac{1}{2}{\cal G}_{ab}(\nabla^c\phi)(\nabla_c\phi)\right]
+\frac{1}{\phi}\Big(\nabla_a\nabla_b\phi-{\cal G}_{ab}\nabla^2\phi\Big)
\end{equation}
and
\begin{equation}\label{(D+1)-equation-2}
\frac{2\omega}{\phi}\nabla^2\phi
-\frac{\omega}{\phi^{^{2}}}{\cal G}^{ab}(\nabla_a\phi)(\nabla_b\phi)+R^{^{(D+1)}}=0,
\end{equation}
respectively, where $\nabla^2\equiv\nabla_a\nabla^a$ and $T^{^{(D+1)}}_{ab}$ is
the energy--momentum tensor (EMT) of the matter fields in
$(D+1)$--dimensional space--time. Contraction of the indices in
Eq.~(\ref{(D+1)-equation-1}) yields
\begin{equation}\label{(D+1)-equation-3}
R^{^{(D+1)}}=-\frac{16\pi\,T^{^{(D+1)}}}{(D-1)\phi}
+\frac{\omega}{\phi^{2}}(\nabla^c\phi)(\nabla_c\phi)
+\frac{2D}{D-1}\frac{\nabla^2\phi}{\phi},
\end{equation}
where $T^{^{(D+1)}}={\cal G}^{ab}T^{^{(D+1)}}_{ab}$. By replacing
(\ref{(D+1)-equation-3}) into (\ref{(D+1)-equation-2}), we
further get
\begin{equation}\label{(D+1)-equation-4}
\nabla^2\phi=\frac{8\pi T^{^{(D+1)}}}{(D-1)\omega+D}.
\end{equation}

In the following, by means of the reduction procedure in the
context of the BD theory, we relate the $(D+1)$--dimensional field
equations to the corresponding ones, with geometrically induced sources,
on the $D$--dimensional space--time. Let us be more precise. We derive the
reduced field equations onto a $D$--dimensional hypersurface by
using the BD Eqs.~(\ref{(D+1)-equation-1}) and
(\ref{(D+1)-equation-4}), in a $(D+1)$--dimensional
space--time described with a line element
\begin{equation}\label{global-metric}
dS^{2}={\cal G}_{ab}(x^c)dx^{a}dx^{b}=
g_{\mu\nu}(x^\alpha,l)dx^{\mu}dx^{\nu}+
\epsilon\psi^2\left(x^\alpha,l\right)dl^{2},
\end{equation}
where the Greek indices run from zero to $(D-1)$, $l$ is a
non--compact coordinate associated to $(D + 1)$th dimension,
which is henceforth labeled with $D$. The indicator
$\epsilon=\pm1$ allows to choose the extra dimension to be
either time--like or space--like, and $\psi$ is a
scalar that depends on all coordinates. Choosing the line
element~(\ref{global-metric}) is obviously restrictive, but it is
also constructive~\cite{FR04}, for, as we will convey, it serves
our herein purposes. We assume that the whole space--time is
foliated by a family of $D$--dimensional hypersurfaces, $\Sigma$,
defined by fixed values of the extra coordinate. Hence, the
intrinsic metric of each hypersurface, e.g. $\Sigma_{0}$ for
$l=l_{0}={\rm constant}$, is obtained by restricting the line
element confined to displacements on it, being orthogonal to the
$(D+1)$--dimensional unit vector
\begin{equation}\label{unitvector}
n^a=\frac{\delta^a_{_D}}{\psi} \qquad {\rm where} \qquad
n_an^a=\epsilon,
\end{equation}
along the extra dimension~\cite{Ponce1,Ponce2}.
Thus, the induced metric $g_{\mu\nu}$ on the hypersurface
$\Sigma_{0}$ has the form
\begin{equation}\label{brane-metric}
ds^{2}={\cal G}_{\mu\nu}(x^{\alpha},
l_{0})dx^{\mu}dx^{\nu}\equiv g_{\mu\nu}dx^{\mu}dx^{\nu}.
\end{equation}
Now, letting $a\rightarrow\mu$ and $b\rightarrow\nu$,
Eq.~(\ref{(D+1)-equation-1}) gives the $D$--dimensional part of
the corresponding $(D+1)$--quantity as
\begin{eqnarray}\label{d+1-Einstein}
G_{\mu\nu}^{^{(D+1)}}\!\!\!&=&\frac{8\pi}{\phi}\,
T_{\mu\nu}^{^{(D+1)}}+\frac{\omega}{\phi^{2}}
\left[({\cal D}_\mu\phi)({\cal D}_\nu\phi)-\frac{1}{2}
g_{\mu\nu}({\cal D}_\alpha\phi)({\cal D}^\alpha\phi)\right]
+\frac{1}{\phi}\left[{\cal D}_\mu{\cal D}_\nu\phi
-g_{\mu\nu}{\cal D}^2\phi\right]\\\nonumber
&-&\!\!\!g_{\mu\nu}\frac{({\cal D}_\alpha\psi)({\cal D}^\alpha\phi)}{\phi\psi}
+\frac{\epsilon g_{\mu\nu,}{}_{_{D}}\phi{}_{_{,D}}}{2\psi^2\phi}-
\frac{\epsilon g_{\mu\nu}}{2\psi^2\phi}\left[2\phi{}_{_{,DD}}+\phi{}_{_{,D}}\left(g^{\alpha\beta}
g_{\alpha\beta,}{}_{_{D}}-\frac{2\psi{}_{_{,D}}}{\psi}+\frac{\omega\phi{}_{_{,D}}}{\phi}\right)\right],
\end{eqnarray}
where ${\cal D}_\alpha$ is the covariant derivative on the
hypersurface, whose computation employs $g_{\mu\nu}$. Furthermore,
the notation $A{}_{_{,D}}$ denotes the derivative of any quantity
$A$ with respect to the extra coordinate $l$, and ${\cal
D}^2\equiv{\cal D}^\alpha{\cal D}_\alpha$. We also have used the
following relations
\begin{eqnarray}\label{rel.1}
(\nabla^c\phi)(\nabla_c\phi)\!\!&=&\!\!({\cal D}^\alpha\phi)({\cal D}_\alpha\phi)+
\epsilon\left(\frac{\phi{}_{_{,D}}}{\psi}\right)^2,\\
\label{rel.2}
\nabla_\mu\nabla_\nu\phi\!\!&=&\!\!{\cal D}_\mu{\cal D}_\nu\phi+
\frac{\epsilon\phi{}_{_{,D}}g_{\mu\nu,}{}_{_{D}}}{2\psi^2},\\
\label{rel.3} \nabla^2\phi\!\!&=&\!\!{\cal D}^2\phi+\frac{({\cal
D}_\alpha\psi)({\cal D}^\alpha\phi)}{\psi}
+\frac{\epsilon}{\psi^2}\left[\phi{}_{_{,DD}}+\phi{}_{_{,D}}
\left(\frac{g^{\mu\nu}g_{\mu\nu,}{}_{_{D}}}{2}-\frac{\psi{}_{_{,D}}}{\psi}\right)\right].
\end{eqnarray}

To obtain the BD effective field equations on the hypersurface, we
should construct the Einstein tensor on the hypersurface.
Therefore, we relate the $R^{^{(D+1)}}_{\alpha\beta}$ and
$R^{^{(D+1)}}$ to their corresponding quantities on the
$D$--dimensional hypersurface. In this respect, we get
\begin{eqnarray}\label{ricci-tensor-D+1,D}
R^{^{(D+1)}}_{\alpha\beta}\!\!\!&=&\!\!\!
R^{^{(D)}}_{\alpha\beta}-\frac{{\cal D}_\alpha{\cal
D}_\beta\psi}{\psi}
+\frac{\epsilon}{2\psi^2}\left[\frac{\psi{}_{_{,D}}g_{\alpha\beta,}{}_{_{D}}}{\psi}
-g_{\alpha\beta,}{}_{_{DD}}-g^{\lambda\mu}g_{\alpha\lambda,}{}_{_{D}}g_{\beta\mu,}
{}_{_{D}}-\frac{1}{2}g^{\mu\nu}g_{\mu\nu,}{}_{_{D}}g_{\alpha\beta,}{}_{_{D}}\right],
\\\nonumber
\\
\label{R_DD}
R^{^{(D+1)}}_{_{DD}}\!\!\!&=&\!\!\!-\epsilon\psi{\cal
D}^2\psi-\frac{1}{4}
g^{\lambda\beta}{}_{_{,D}}g_{\lambda\beta,}{}_{_{D}}
-\frac{1}{2}g^{\lambda\beta}g_{\lambda\beta,}{}_{_{DD}}
+\frac{1}{2\psi}g^{\lambda\beta}g_{\lambda\beta,}{}_{_{D}}\psi{}_{_{,D}}.
\end{eqnarray}
Also, by using Eqs.~(\ref{(D+1)-equation-1}),
(\ref{(D+1)-equation-3}), (\ref{(D+1)-equation-4}) and
(\ref{R_DD}), we obtain
\begin{eqnarray}\label{D2say}
\frac{{\cal D}^2\psi}{\psi}=\!\!\!\!\!&&-\frac{({\cal
D}_\alpha\psi)({\cal D}^\alpha\phi)}{\psi\phi}
-\frac{\epsilon}{2\psi^2}
\left[g^{\lambda\beta}g_{\lambda\beta,}{}_{_{DD}}+\frac{1}{2}g^{\lambda\beta}{}_{_{,D}}
g_{\lambda\beta,}{}_{_{D}}-\frac{g^{\lambda\beta}g_{\lambda\beta,}{}_{_{D}}\psi{}_{_{,D}}}{\psi}\right]\\\nonumber
\!\!\!\!\!&&-\frac{\epsilon}{\psi^2\phi}
\left[\phi{}_{_{,DD}}+\phi{}_{_{,D}}\left(\frac{\omega\phi{}_{_{,D}}}{\phi}
-\frac{\psi{}_{_{,D}}}{\psi}\right)\right]
+\frac{8\pi}{\phi}\left[\frac{(\omega+1)T^{^{(D+1)}}}{(D-1)\omega+D}
-\frac{\epsilon T^{^{(D+1)}}_{_{DD}}}{\psi^2}\right],
\end{eqnarray}
where the relation
$g^{\mu\beta}g^{\lambda\sigma}g_{\lambda\beta,}{}_{_{D}}
g_{\mu\sigma,}{}_{_{D}}+g^{\mu\sigma}{}_{_{,D}}g_{\mu\sigma,}{}_{_{D}}=0$
has been used. By applying relations~(\ref{R_DD}) and
(\ref{D2say}), we can relate the Ricci scalar in
$(D+1)$--dimensional space--time to its corresponding one on the
hypersurface, as
\begin{eqnarray}\label{R(D+1)-R(D)}
R^{^{(D+1)}}=\!\!\!\!\!&&R^{^{(D)}}+\frac{2({\cal D}_\alpha\psi)
({\cal D}^\alpha\phi)}{\psi\phi}
-\frac{\epsilon}{4\psi^2}\left[g^{\alpha\beta}{}_{_{,D}}
g_{\alpha\beta,}{}_{_{D}}+\left(g^{\alpha\beta}g_{\alpha\beta,}{}_{_{D}}\right)^2\right]\\\nonumber
\!\!\!\!\!&&+\frac{2\epsilon}{\psi^2\phi}\left[\phi{}_{_{,DD}}+\phi{}_{_{,D}}
\left(\frac{\omega\phi{}_{_{,D}}}{\phi}-\frac{\psi{}_{_{,D}}}{\psi}\right)\right]
+\frac{16\pi}{\phi}\left[\frac{\epsilon T^{^{(D+1)}}_{_{DD}}}{\psi^2}
-\frac{(\omega+1)T^{^{(D+1)}}}{(D-1)\omega+D}\right].
\end{eqnarray}
By using the above expressions, we can eventually obtain the
reduced equations onto the $D$--dimensional hypersurface.
 This will produce our $D$--dimensional MBDT scenario.
In what follows, we outline these retrieved equations in
three separated steps, providing suitable interpretations.

Firstly, by applying
equations~(\ref{d+1-Einstein}), (\ref{ricci-tensor-D+1,D})
and (\ref{R(D+1)-R(D)}), we construct the Einstein equations on the hypersurface as
\begin{eqnarray}\label{BD-Eq-DD}
G_{\mu\nu}^{^{(D)}}\!\!&=&\!\!\frac{8\pi}{\phi}\,
\left(S_{\mu\nu}+T_{\mu\nu}^{^{[\rm BD]}}\right)+
\frac{\omega}{\phi^2}\left[({\cal D}_\mu\phi)({\cal D}_\nu\phi)-
\frac{1}{2}g_{\mu\nu}({\cal D}_\alpha\phi)({\cal
D}^\alpha\phi)\right] \cr
 &&+\frac{1}{\phi}\left[{\cal D}_\mu{\cal
D}_\nu\phi- g_{\mu\nu}{\cal
D}^2\phi\right]-g_{\mu\nu}\frac{V(\phi)}{2\phi} \cr
 & \equiv &
\frac{8\pi}{\phi} T_{\mu\nu}^{^{(D)[{\rm eff}]}} +
\frac{\omega}{\phi^2}\left[({\cal D}_\mu\phi)({\cal D}_\nu\phi)-
\frac{1}{2}g_{\mu\nu}({\cal D}_\alpha\phi)({\cal
D}^\alpha\phi)\right] + \frac{1}{\phi}\left[{\cal D}_\mu{\cal
D}_\nu\phi- g_{\mu\nu}{\cal D}^2\phi\right]-g_{\mu\nu}\frac{V(\phi)}{2\phi}.
\end{eqnarray}
The above result conveys the standard BD equations that contain an
induced scalar potential, though, there are a few points which we
should make clear:
\begin{itemize}
\item
    $S_{\mu\nu}$ represents the effects of the $(D+1)$--dimensional EMT on
    the hypersurface and is given by
\begin{eqnarray}\label{S}
S_{\mu\nu}\equiv T_{\mu\nu}^{^{(D+1)}}-
g_{\mu\nu}\left[\frac{(\omega+1)T^{^{(D+1)}}}{(D-1)\omega+D}-
\frac{\epsilon\, T_{_{DD}}^{^{(D+1)}}}{\psi^2}\right].
\end{eqnarray}
 Clearly, if one assumes that the $(D+1)$--dimensional space--time is empty
 of the usual matter fields [i.e., no $L_{_{\rm matt}}^{^{(D+1)}}$ term in action (2.1)],
 then $S_{\mu\nu}$ will vanish.

\item
  The quantity $T_{\mu\nu}^{^{[\rm BD]}}$  is an induced EMT
for a BD theory in $D$ dimensions and, in turn, it contains three
components, namely,
\begin{eqnarray}\label{matt.def}
T_{\mu\nu}^{^{[\rm BD]}}= T_{\mu\nu}^{^{[\rm IMT]}}+T_{\mu\nu}^{^{[\rm \phi]}}
+\frac{1}{16\pi}g_{\mu\nu}V(\phi),
\end{eqnarray}
where
\begin{eqnarray}\label{IMTmatt.def}
\frac{8\pi}{\phi}T_{\mu\nu}^{^{[\rm IMT]}}&\!\!\!\equiv &\!\!
\frac{{\cal D}_\mu{\cal D}_\nu\psi}{\psi}
-\frac{\epsilon}{2\psi^{2}}\left(\frac{\psi_{_{,D}}
g_{\mu\nu,}{}_{_{D}}}{\psi}-g_{\mu\nu,}{}_{_{DD}}
+g^{\lambda\alpha}g_{\mu\lambda,}{}_{_{D}}g_{\nu\alpha,}
{}_{_{D}}-\frac{1}{2}g^{\alpha\beta}g_{\alpha\beta,}
{}_{_{D}}g_{\mu\nu,}{}_{_{D}}\right)\cr
 \!\!\!&&\!\!-\frac{\epsilon g_{\mu\nu}}{8\psi^2}
\left[g^{\alpha\beta}{}_{_{,D}}g_{\alpha\beta,}{}_{_{D}}
+\left(g^{\alpha\beta}g_{\alpha\beta,}{}_{_{D}}\right)^{2}\right],\\\nonumber
\\
\label{T-phi} \frac{8\pi}{\phi}T_{\mu\nu}^{^{[\rm
\phi]}}&\!\!\!\equiv &\!\!
\frac{\epsilon\phi_{_{,D}}}{2\psi^2\phi}\left[g_{\mu\nu,}{}_{_{D}}
+g_{\mu\nu}\left(\frac{\omega\phi_{_{,D}}}{\phi}-g^{\alpha\beta}g_{\alpha\beta,}
{}_{_{D}}\right)\right].
\end{eqnarray}
The first part of the induced EMT, i.e. $T_{\mu\nu}^{^{[\rm
IMT]}}$, is the $(D+1)$th part of the metric (\ref{global-metric}) which is
geometrically induced on the hypersurface. In fact, as the BD scalar field
plays (inversely) the role of the Newton gravitational constant,
we can deduce that this part is the modified version of the
induced EMT, introduced in the IMT scenario. Whereas, the second
part, i.e. $T_{\mu\nu}^{^{[\rm \phi]}}$, depends on the BD scalar
field and its derivatives with respect to the $(D+1)$th coordinate, has no analogue in IMT.

\item
  The quantity introduced by $V(\phi)$ is the induced scalar potential
  on the hypersurface, which is derived from the other
  reduced equation on the hypersurface, see Eq.~(\ref{v-def}).
\end{itemize}

Secondly, we obtain the $D$--dimensional counterpart of
Eq.~(\ref{(D+1)-equation-4}), the wave equation on the
hypersurface. By contracting Eq.~(\ref{BD-Eq-DD}), we get a
relation between $R^{^{(D)}}$, $S=g^{\mu\nu}S_{\mu\nu}$ and
$T^{^{[\rm BD]}}=g^{\mu\nu}T_{\mu\nu}^{^{[\rm BD]}}$ as
\begin{eqnarray}\label{R-D}
R^{^{(D)}}=-\frac{16\pi}{(D-2)\phi}\left(S+T^{^{[\rm BD]}}\right)
+\frac{\omega({\cal D}_\alpha\phi)({\cal D}^\alpha\phi)}{\phi^2}+
\frac{2(D-1)}{(D-2)}\frac{{\cal D}^2\phi}{\phi}
+\frac{DV(\phi)}{(D-2)\phi}.
\end{eqnarray}
Then, by substituting relations~(\ref{R(D+1)-R(D)})
and~(\ref{R-D}) into Eq.~(\ref{(D+1)-equation-2}) and applying
relations~(\ref{rel.2}) and (\ref{rel.3}), we finally achieve
\begin{eqnarray}\label{D2-phi}
{\cal D}^2\phi=\frac{8\pi}{(D-2)\omega+(D-1)}\left(S+T^{^{[\rm BD]}}\right)+
\frac{1}{(D-2)\omega+(D-1)}\left[\phi\frac{dV(\phi)}{d\phi}-\frac{D}{2}V(\phi)\right],
\end{eqnarray}
where
\begin{eqnarray}\label{v-def}
\phi \frac{dV(\phi)}{d\phi}\!\!&\equiv&\!\!-(D-2)(\omega+1)
\left[\frac{({\cal D}_\alpha\psi)({\cal D}^\alpha\phi)}{\psi}
+\frac{\epsilon}{\psi^2}\left(\phi_{_{,DD}}-
\frac{\psi_{_{,D}}\phi_{_{,D}}}{\psi}\right)\right]\\\nonumber
&&\!\!\!-\frac{(D-2)\epsilon\omega\phi_{_{,D}}}{2\psi^2}
\left[\frac{\phi_{_{,D}}}{\phi}+g^{\mu\nu}g_{\mu\nu,}{}_{_{D}}\right]
+\frac{(D-2)\epsilon\phi}{8\psi^2}
\Big[g^{\alpha\beta}{}_{_{,D}}g_{\alpha\beta,}{}_{_{D}}
+(g^{\alpha\beta}g_{\alpha\beta,}{}_{_{D}})^2\Big]\\\nonumber
&&\!\!\!+8\pi(D-2)\left[\frac{(\omega+1)T^{^{(D+1)}}}{(D-1)
\omega+D}-\frac{\epsilon\, T_{_{DD}}^{^{(D+1)}}}{\psi^2}\right].
\end{eqnarray}
Hence, in this applied approach, the dimensional reduction procedure provides
an expression to obtain the potential, up to a constant
of integration, rather than being merely introduced by hand.

Finally, we derive the counterpart equation for a conservation
equation introduced within the IMT. For this purpose, by
substituting $a\rightarrow\alpha$ and $b\rightarrow D$ in Eq.
(\ref{(D+1)-equation-1}), we get
\begin{eqnarray}\label{G-D,alpha}
G_{\alpha D}^{^{(D+1)}}=R_{\alpha D}^{^{(D+1)}}=\frac{8\pi}{\phi}
T^{^{(D+1)}}_{\alpha D}+\frac{\omega\phi_{_{,D}}}{\phi^2}({\cal
D}_\alpha\phi) +\frac{1}{\phi}{\cal
D}_\alpha\left(\phi_{_{,D}}\right)
-\frac{1}{2\phi}g_{\alpha\beta,}{}_{_{D}}({\cal D}^\beta\phi)-
\frac{\phi{}_{_{,D}}({\cal D}_\alpha\psi)}{\psi\phi},
\end{eqnarray}
where the first equality comes from the metric~(\ref{global-metric}).
On the other hand, metric (\ref{global-metric}) for the mentioned
component gives
\begin{equation}\label{GDmo-2}
G^{^{(D+1)}}_{\alpha D}= \psi P^{\beta}{}_{\alpha;\beta},
\end{equation}
where $P_{\alpha\beta}$ is given by
\begin{equation}\label{P-mono}
P_{\alpha\beta}\equiv\frac{1}{2
\psi}\left(g_{\alpha\beta,}{}_{_{D}}
-g_{\alpha\beta}g^{\mu\nu}g_{\mu\nu,}{}_{_{D}}\right).
\end{equation}
Therefore, Eqs.~(\ref{G-D,alpha}) and (\ref{GDmo-2})
give the dynamical equation for $P_{\alpha\beta}$ as
\begin{eqnarray}\label{P-Dynamic}
P^{\beta}{}_{\alpha;\beta}&=&\!\!
\frac{8\pi}{\psi \phi}
T^{^{(D+1)}}_{\alpha D}+\frac{\omega\phi_{_{,D}}}{\psi\phi^2}\left({\cal D}_\alpha\phi\right)
+\frac{1}{\psi\phi}{\cal D}_\alpha\left(\phi_{_{,D}}\right)\cr
&-&\!\!\!\frac{1}{2\psi \phi}g_{\alpha\lambda,}{}_{_{D}}\left({\cal D}^\lambda\phi\right)-
\frac{\phi{}_{_{,D}}({\cal D}_\alpha\psi)}{\psi^2\phi}\,.
\end{eqnarray}

As we conclude this section, let us further clarify a few points
about the herein retrieved $D$--dimensional MBDT.
\begin{itemize}
\item
 The $(D+1)$--dimensional field
equations (\ref{(D+1)-equation-1}) and (\ref{(D+1)-equation-4}),
with a general metric (\ref{global-metric}), split naturally into
four sets of Eqs. (\ref{D2say}), (\ref{BD-Eq-DD}), (\ref{D2-phi})
and (\ref{P-Dynamic}). As mentioned, Eqs. (\ref{BD-Eq-DD}) and
(\ref{D2-phi}) are the BD field equations on a $D$--dimensional
space--time, with a geometrically induced energy--momentum
source\rlap.\footnote{More precisely, they are retrieved from the
action
 ${\cal S}^{^{(D)}}\!\!\!\!
=\!\int d^{^{\,D}}\!x \sqrt{-g}\,\left[\phi
R^{^{(D)}}-\frac{\omega}{\phi}\, g^{\alpha\beta}\,({\cal
D}_\alpha\phi)({\cal D}_\beta\phi)-V(\phi)+16\pi\,
L\!^{^{(D)}}_{_{\rm matt}}\right]$,
 where specifically
$\sqrt{-g}\left(S_{\alpha\beta}+T^{^{[\rm
BD]}}_{\alpha\beta}\right)\equiv 2\delta\left( \sqrt{-g}\,
L\!^{^{(D)}}_{_{\rm matt}}\right)/\delta g^{\alpha\beta}$.}\
 Such a correspondence is guaranteed
by the Campbell--Magaard theorem~\cite{C26,M63,RTZ95,LRTR97,SW03}.
Furthermore, it is important to note that Eq.~(\ref{D2say}) has no
 standard BD analog, and the set of Eqs.~(\ref{P-Dynamic}) is a
generalized conservation law introduced within the IMT.
\item
 The induced EMT is covariantly conserved
(the same way as in the standard four--dimensional BD theory),
i.e. ${\cal D}_\beta T^{^{[\rm BD]}}_\alpha{}^{\beta}=0$.
\item
  In the particular case of $D=4$ (and $L\!^{^{(D+1)}}_{_{\rm
matt}}=0$), the MBDT approach reproduces the results of~\cite{Ponce2} (and~\cite{Ponce1}). In addition, when
$L\!^{^{(D+1)}}_{_{\rm matt}}=0$ and the BD scalar field takes
constant values, the results of~\cite{RRT95} are also reproduced,
as expected.
\item
  In the special case\footnote{The case $\omega=-1$ corresponds
  precisely to the value predicted when the BD theory is derived
as the low energy limit of some string
theories~\cite{Faraoni.book,BD12}.}\
 when $\omega=-1$, $l$ is a
cyclic coordinate and $L\!^{^{(D+1)}}_{_{\rm matt}}=0$, the scalar
potential, without loss of generality, vanishes. Thus, to
reproduce a general version of a $D$--dimensional BD theory by means
of the above dimensional reduction procedure, we should notice those
requirements\rlap.\footnote{Also, we should notice that when the
coupling parameter $\omega$ goes to infinity, with suitable
boundary conditions, the approach developed in this section may be
viewed as a generalization of the procedure of~\cite{RRT95}
(but not always, corresponding to the content in~\cite{BR93,BS97,Faraoni99}).}
\item
 In the case where $T^{^{(D+1)}}_{\alpha D}=0$ and the BD scalar
field takes constant values, Eq. (\ref{P-Dynamic}) reduces to
$P^{^{(D)\beta}}{}_{\alpha;\beta}=0$. In the brane world theory, this
quantity is proportional to the EMT of the matter on the
brane~\cite{Ponce2,Pon01}. If, in addition, $l$ is a cyclic
coordinate then, equation (\ref{P-Dynamic}) will reduce to an
identity.
\end{itemize}

\indent

We would like to close this section by indicating an interesting point
regarding how the BD theory can be related to the KK setting.
An important benefit of the MBDT is that the induced matter and scalar potential,
which depend on the BD scalar field and its derivatives,
derived via the BD action (\ref{(D+1)-action}),
 should be regarded as
 fundamental quantities rather than quantities added by hand.
 However, some questions may be asked:
can we accept the BD scalar field in action
(\ref{(D+1)-action}) as a fundamental field?
 Where does it emerge from?
%Indeed, the gravitational sector of the Einstein-Hilbert action
% in GR is purely geometric, but in the original
 %BD theory (in Jordan frame), the BD scalar field, which decoupled from the
% matter part, is responsible to describe the
 %gravitational field together with the metric tensor, may not sound
% to have geometrical origin~\cite{scalarbook,Faraoni.book}.}

 In Appendix~\ref{App.A}, by generalizing the approaches of~\cite{scalarbook, PS02, Faraoni.book}, we show that
  the $(D+1)$--dimensional BD framework (in vacuum)
  %by accepting the KK three key assumptions,
  can be derived from a generalized GR (i.e., a simplest version of the KK) theory in
  a $(D+1+d)$--dimensional space--time by obtaining
  $\omega=-1+1/d$ in which $d$ is the number of the compactified extra
  spatial dimensions. Moreover, in this formalism, the BD scalar field emerges as a
  geometrical quantity, namely, it is related to the determinant of the metric
  associated to submanifold of extra dimensions.

  %We should note that the results of
  %Appendix~\ref{App.A} never have been used in the other parts of this paper.}

  %\bl{Another appropriate answer to the above questions can be
  %supplied by applying the Lyra geometry~\cite{Lyra},
 % by which, it has been shown that the gravitational sector of
   %the BD theory only contains geometrical terms~\cite{Faraoni.book, Sol88}.}

%\textcolor[rgb]{0.00,0.00,1.00}{$
%\clubsuit\diamondsuit\heartsuit\spadesuit $} In the next two
%sections, we will investigate cosmological applications of this
%section features. The following case s studies proposed to
%illustrate the procedure herein constructed. We should remind that,
%in order to avoid the so-called ghost fields in a $D$--dimensional
%space--time, we will assume $\omega>-(D-1)/(D-2)$. Moreover, when the
%coupling parameter $\omega$ goes to infinity, with suitable boundary
%conditions, the approach developed in this section may be viewed as
%a generalization of the procedure of~\cite{RRT95} (but not
%always, corresponding to the content in~\cite{BR93,BS97,Faraoni99}).
%\textcolor[rgb]{0.00,0.00,1.00}{$
%\clubsuit\diamondsuit\heartsuit\spadesuit $}
\section{exact solutions of BD cosmology in ($D+1$)--Dimensional vacuum space--time}
\label{OT-solution}

\indent

In this section, we assume an
empty\footnote{Eqs.~(\ref{(D+1)-equation-1}) and
(\ref{(D+1)-equation-4}) are the field equations of the BD theory
in $(D+1)$--dimensions, though (as described in the Introduction),
we propose to employ them in the this section, in terms of
``vacuum'' cosmological solutions, which are defined as a
configuration where there is no other matter source (except the BD
scalar field) in $(D+1)$--dimensional space--time. In this case,
Eq.~(\ref{(D+1)-equation-2}) becomes
\begin{equation}\label{vacuum ricci}\nonumber
R^{^{(D+1)}}=\frac{\omega}{\phi^{2}}{\cal
G}^{ab}(\nabla_a\phi)(\nabla_b\phi),
\end{equation}
which means that the $(D+1)$--dimensional scalar curvature is
generated only by a free scalar field~\cite{ARB07}.}\
 $(D+1)$--dimensional
space--time that is described by an extended version of
the~Friedmann--Lema\^{\i}tre--Robertson--Walker~(FLRW)
metric
%\footnote{Note that this line--element is indeed
%the $(D+1)$--dimensional analogue of the one used in~\cite{Ponce1}.
%Thus, if we assume $D=4$, some results from
%~\cite{Ponce1} and~\cite{Ponce2} will be reproduced.}\
 and then, in section~(\ref{OT-reduced}), by means of
the MBDT procedure described in the previous section, we
investigate the cosmology reduced on a $D$--dimensional
hypersuface. Furthermore, in order to respect the space--time
symmetries, we assume that the metric components
 and the BD scalar field depend only on the
comoving time\footnote{In~\cite{RFS11}, it has been
shown that, in a five--dimensional space--time, when the usual three scale factors are functions of
 the cosmic time whereas the scale factor of the extra dimension is a constant (i.e. $\psi={\rm constant}$),
if the BD scalar field is assumed as a function of both $t$ and $l$,
%the cosmic time \emph{and} the extra coordinate,
then, in general, we will encounter inconsistencies in
the field equations.}. Moreover, we choose the $K=0$
case with the metric
%\footnote{The analyzed astrophysical data, obtained from
%WMAP~\cite{Spe03} and BOOMERANG~\cite{Ber00} based on GR, shows
%that $k=0$ around the present epoch. Notwithstanding, any other
%gravitational theory may present a different
%answer~\cite{Ponce1}.}
\begin{equation}\label{ohanlon metric}
dS^{2}=-dt^{2}+a^{2}(t)\left[\sum^{D-1}_{i=1}
\left(dx^{i}\right)^{2}\right]+\epsilon \psi^2(t)dl^{2},
\end{equation}
where $t$ is the cosmic time, $a(t)$ and $\psi(t)$ are
cosmological scale factors and $x^i$'s are the Cartesian coordinates.

The dynamical field equations in
vacuum\footnote{In the present work, we leave a few
more extended solutions that can be
produced by assuming the following general cases: i) taking the BD scalar field and metric components
such that
%not only do they depend on the cosmic time, but
they also depend on the
spatial coordinates, specially, the extra coordinate $l$,
 ii) assuming an ordinary matter in the bulk, iii) $K\neq0$ and/or iv) considering a more
flexible embedding approach~\cite{Leon06, Leon06-1, Leon09}.
Considering such assumptions would make the analysis more realistic.
%, but, we should note that to
%obtain the exact analytical solutions we have to impose a few {\it ansatzs} to simplify the calculations.
 %Concerning the assumption ii),
Concerning the second assumption, we should stress that, in this work, we have
 been studying the BD theory in the Jordan
frame in a $(D+1)$--dimensional space-time, in which the $\phi$ is seen as a (scalar)
part of the gravitational degrees of freedom rather than a matter degree of freedom
(where it could play the role of a $k$--essence
field,~see, e.g.,~\cite{Kim04,Kim05} in a $4$--dimensional space--time).
Moreover,
%from now on, following the Kaluza's first key assumption,
we will assume henceforth
that there is no ordinary matter in $(D+1)$--dimensional
space--time. These assumptions
%assumptions lead us to the
allow to extract the induced matter and the scalar potential as a
manifestation of pure geometry in a $(D+1)$--dimensional world~\cite{OW97,stm99}.
Within this context, the suggestion is to replace the
``base wood'' of matter by the ``pure marble" of geometry~\cite{DNP86}.} are given by
\begin{eqnarray}\label{ohanlon-eq-1}
\ddot{\phi}+\dot{\phi}\left[(D-1)\frac{\dot{a}}{a}+\frac{\dot{\psi}}{\psi}\right]\!\!&=&\!\!0,\\
\label{ohanlon-eq-2}
(D-1)\frac{\dot{a}}{a}\left(\frac{D-2}{2}\frac{\dot{a}}{a}+\frac{\dot{\psi}}{\psi}\right)
\!\!&=&\!\!\frac{1}{\phi}\left(\ddot{\phi}+\frac{\omega}{2}\frac{\dot{\phi}^2}{\phi}\right),\\
\label{ohanlon-eq-3}
(D-2)\frac{\ddot{a}}{a}+(D-2)\frac{\dot{a}}{a}
\left[\frac{D-3}{2}\frac{\dot{a}}{a}+\frac{\dot{\psi}}{\psi}\right]
+\frac{\ddot{\psi}}{\psi}\!\!&=&\!\!\frac{\dot{\phi}}{\phi}
\left(\frac{\dot{a}}{a}-\frac{\omega}{2}\frac{\dot{\phi}}{\phi}\right),\\
\label{ohanlon-eq-4}
(D-1)\left[\frac{\ddot{a}}{a}+\frac{D-2}{2}\left(\frac{\dot{a}}{a}\right)^2\right]
\!\!&=&\!\!\frac{\dot{\phi}}{\phi}\left(\frac{\dot{\psi}}{\psi}-\frac{\omega}{2}\frac{\dot{\phi}}{\phi}\right),
\end{eqnarray}
where equation~(\ref{ohanlon-eq-1}) is obtained
from~(\ref{(D+1)-equation-4}), and equations~(\ref{ohanlon-eq-2}),
(\ref{ohanlon-eq-3}) and (\ref{ohanlon-eq-4}) are associated to
the components $a=0=b$, $a=b=1,2,\cdots,(D-1)$ and $a=D=b$ of
Eq.(\ref{(D+1)-equation-1}), respectively, in which we have used
 equation~(\ref{ohanlon-eq-1}). ``\,\,${\bf\dot{}}$\,\,'' denotes the derivative
with respect to the cosmic time.

Let us solve Eqs.~(\ref{ohanlon-eq-1})-(\ref{ohanlon-eq-4}) by
using the power--law solutions\footnote{The power--law solutions, in the conventional
BD theory, have resemblance to the inflationary
de~Sitter attractor in GR. However, in the scalar-tensor gravity, these solutions
have been assumed for investigating the quintessence models~\cite{Faraoni.book}.}
%\footnote{In four--dimensional BD theory, such
%solutions play a role analogous to that of the inflationary
%de~Sitter attractor in GR. When investigating quintessence models
%(a non--vacuum situation), these types of solutions are assumed,
%in which the exponents $r$ and $s$ are determined {\it a
%posteriori}, together with the other
%parameters~\cite{Faraoni.book}.}
\begin{equation}\label{power-law-phi}
a(t)=a_{0}\left(\frac{t}{t_{0}}\right)^{r}, \hspace{10mm}
\psi(t)=\psi_{0}\left(\frac{t}{t_{0}}\right)^{n}\hspace{5mm}
{\rm and} \hspace{5mm}  \phi(t)=\phi_{0}\left(\frac{t}{t_{0}}\right)^{s}\,,
\end{equation}
where $a_{0}$, $\psi_{0}$ and $\phi_{0}$ are constants determined
in an arbitrary fixed time $t_0$, and $r$, $n$ and $s$ are
parameters, which are not independent,
satisfying the field equations. Substituting these
solutions in equations~(\ref{ohanlon-eq-1})--(\ref{ohanlon-eq-4}),
it yields
\begin{equation}\label{s-om}
s=1-(D-1)r-n \hspace{10mm}
{\rm and} \hspace{10mm} \omega=-\frac{D(D-1)r^2+2(n-1)\Big[(D-1)r+n\Big]}{\Big[1-(D-1)r-n\Big]^2},
\end{equation}
where, by assuming negative values for $n$, we must have
\begin{equation}\label{rest.on.omega}
\omega\geq-\frac{n(D+1)+(D-1)}{nD+(D-2)} \hspace{10mm}
{\rm and} \hspace{10mm}
\omega\neq0,-D/(D-1).
\end{equation}

%We should mention that all the resulted ranges for the
%BD coupling parameter must satisfy the above constraints.}
%In the rest of this section, we would like to restrict ourselves to investigate some especial solutions.
Some special solutions, that are of our interest, include:
\begin{itemize}
\item
 When $s$ tends to zero, then $\omega$ goes to infinity and $\phi$
takes a constant value. In this limit, the field equations are
only satisfied\footnote{We disregard the static case of $r=0$.}\
 for $r=2/D$, i.e. $n=-(D-2)/D$. Thus, the
$(D+1)$--dimensional solution
\begin{equation}\label{GR-ohanlon metric}
dS^{2}=-dt^{2}+a_0^{2}\left(\frac{t}{t_0}\right)^{\frac{4}{D}}
\left[\sum^{D-1}_{i=1}\left(dx^{i}\right)^{2}\right]
+\epsilon\psi_0^2\left(\frac{t}{t_0}\right)^{-2(D-2)/D}dl^{2}
\end{equation}
is obtained, which is the unique solution for the
$(D+1)$--dimensional metric~(\ref{ohanlon metric}) for the
Einstein field equations in vacuum.

\item
 In the case where $\omega=-D/(D-1)$, we cannot set this value of
$\omega$ in solutions (\ref{s-om}) and then get the values for the
exponents $r$, $n$ and $s$. However, instead, we should start from
the field equations~(\ref{ohanlon-eq-1})--(\ref{ohanlon-eq-4}). It
is straightforward to show that for $\omega=-D/(D-1)$, we have a
$(D+1)$--dimensional de~Sitter--like space
\begin{equation}\label{s2}
a(t)=\psi(t)=a'_{0}e^{\xi t}  \qquad {\rm and} \qquad
\phi(t)=\phi'_{0}e^{-D\xi t}\,,
\end{equation}
where $\xi$, $a'_{0}$ and $\phi'_{0}$ are constants.
\item
 One of the most well--known class of solutions in standard BD
theory is the O'Hanlon and Tupper
solution~\cite{o'hanlon-tupper-72}.
This class
corresponds to ``vacuum'' with a free scalar and the
range of the BD parameter in four--dimensional space--time is
restricted to $\omega>-3/2$, $\omega\neq0, -4/3$~\cite{Faraoni.book}.
%More precisely, the O'Hanlon and Tupper solution describes an empty universe, the
%curvature of which is produced by a non--static scalar field, that
%is usually related to the Newtonian constant of gravitation
%through $\phi\sim G^{-1}$.
Assuming $a(t)=\psi(t)=a_{0}\left(t/t_{0}\right)^{r}$ and $\omega\neq-D/(D-1)$ at
the beginning, thus solutions~(\ref{power-law-phi}) and
(\ref{s-om}) are reduced to a generalized O'Hanlon and Tupper solution in a
($D+1$)--dimensional space--time as
\begin{equation}\label{ohanlon-metric-particular}
dS^{2}=-dt^{2}+a_0^{2}\left(\frac{t}{t_0}\right)^{2r}
\left[\sum^{D-1}_{i=1}\left(dx^{i}\right)^{2}+dl^{2}\right], \hspace{10mm}
\phi(t)=\phi_{0}\left(\frac{t}{t_0}\right)^{s}
\end{equation}
where
\begin{equation}\label{q}
r_{\pm}=\frac{1}{(D+1)+D\omega}
\left[\omega+1\pm\sqrt{\frac{(D-1)\omega+D}{D}}\right]
\end{equation}
and
\begin{equation}\label{s}
s_{\pm}=\frac{1\mp\sqrt{D}\sqrt{D+(D-1)\omega}}{(D+1)+D\omega}\,,
\end{equation}
where $\omega>-D/(D-1)$ and $r_{\pm}$ and $s_{\pm}$ algebraically are
related by constraint\footnote{For convenience, we will drop the
index $\pm$ from the parameters $r$ and $s$.}\
$s+Dr=1$. We also assumed $\epsilon=1$. When the cosmic time goes to zero, this
solution has a big bang singularity. In a four--dimensional
space--time, this solution has been obtained by means of different
methods~\cite{o'hanlon-tupper-72,KE95,MW95,Faraoni.book}. We can
easily show that
\begin{equation}\label{omeg2}
\omega=\frac{1-s}{Ds^2}\Big[(D+1)s+(D-1)\Big].
\end{equation}
When $s$ tends to zero, then $\omega$ goes to
infinity and $\phi$ takes a constant value. In this limit, we have
\begin{equation}\label{GR-ohanlon metric}
dS^{2}=-dt^{2}+a_0^{2}t^{\frac{2}{D}}
\left[\sum^{D-1}_{i=1}\left(dx^{i}\right)^{2}+dl^{2}\right].
\end{equation}
We should note that in the mentioned limit, the corresponding
general relativistic solution is~not
reproduced; as~(\ref{GR-ohanlon metric}) illustrates, it is not a Minkowski space.
Let us check it for, e.g., a four--dimensional
space--time (i.e. by setting $D=3$). In this special case, the
solutions are reduced to
\begin{equation}\label{omega limit}
a(t)=a_0\left(\frac{t}{t_0}\right)^{1/3}
\qquad {\rm and}\qquad \phi={\rm constant},
\end{equation}
where the scale factor has a decelerated expanding
behavior~\cite{Faraoni.book,RFK11}.
\end{itemize}

In the next section, as an application of the MBDT in cosmology,
we proceed to investigate the effective $D$--dimensional picture
generated by the exact power--law solutions in $(D+1)$--dimensional
space--time.

\section{Reduced Brans--Dicke Cosmology in $D$ Dimensions}
\label{OT-reduced}
\indent

The non--vanishing components of the induced EMT (\ref{matt.def})
associated to the metric~(\ref{ohanlon metric}) on the
hypersurface $\Sigma_0$ are
\begin{eqnarray}\label{t-00}
\frac{8\pi}{\phi}T^{0[{\rm BD}]}_{\,\,\,0}\!\!\!&=&\!\!\!
-\frac{\ddot{\psi}}{\psi}+\frac{V(\phi)}{2\phi},\\\nonumber
\\
\label{t-ii}
\frac{8\pi}{\phi}T^{i[{\rm BD}]}_{\,\,\,i}\!\!\!&=&\!\!\!
-\frac{\dot{a}\dot{\psi}}{a\psi}+\frac{V(\phi)}{2\phi},
\end{eqnarray}
where the induced potential $V(\phi)$ will be determined from
(\ref{v-def}). As the different components of $T^{i[{\rm BD}]}_{\,\,\,i}$
[where $i=1,2,\cdots,(D-1)$ with no sum] are
equal, thus the induced--matter can be considered as a perfect
fluid with an energy density $\rho_{_{\rm BD}}\equiv -T^{0[{\rm BD}]}_{\,\,\,0}$
and isotropic pressures $p_{_{\rm BD}}=p_i\equiv T^{i[{\rm BD}]}_{\,\,\,i}$.

In order to derive the induced scalar potential,
we substitute the power--law solutions (\ref{power-law-phi})
into (\ref{v-def}) and evaluate it on the hypersurface. Thus, we get
\begin{equation}\label{Eq-pot}
\phi\frac{dV}{d\phi}{\Biggr|}_{_{\Sigma_{o}}}\!\!\!\!\!=
(D-2)(\omega+1)nst_0^{-2}\phi_0^{2/s}\phi^{\frac{s-2}{s}},
\end{equation}
where $t_0$, $\phi_0$, $n$, $s$ and $\omega$ are given
by relations~(\ref{power-law-phi}) and (\ref{s-om}).
By integrating this equation, we obtain
\begin{equation}\label{pot.gen}
V(\phi)=\left \{
 \begin{array}{c}
 \frac{2\phi_0}{t_0^2}(D-2)n
 \left(\omega+1\right)
 {\rm ln}\phi
 \hspace{20mm} {\rm for}\hspace{5mm} s=2\\\nonumber \\
 \frac{\phi_0^{2/s}ns^2}{t_0^2(s-2)}(D-2)(\omega+1)
\phi^{(s-2)/s}
  \hspace{15mm} {\rm for}\hspace{5mm} s\neq0,2,\\
 \end{array}\right.
\end{equation}

where the constants of integration have been set equal to zero. In
the special cases, regardless of $D$, where $\omega=-1$ or $n=0$,
the scalar potential will be zero.
Also, for the particular case of $s=0$ where the BD scalar field takes constant values,
it is straightforward to show that
$\frac{dV}{d\phi}{\Biggr|}_{_{\Sigma_{o}}}$ identically vanishes, and thus,
without loss of generality, we can set $V=0$ in this case. From now on, we will~not
investigate the logarithmic potential with $s=2$, for it leads to
some difficulties when the weak energy condition is applied.

By substituting the scalar potential (for $s\neq0,2$) and also
the power--law solutions~(\ref{power-law-phi}) into
relations~(\ref{t-00}) and~(\ref{t-ii}), we get the induced
quantities in a $D$--dimensional hypersurface as
\begin{equation}\label{ro-BD}
\rho_{_{\rm BD}}=\frac{\phi_0}{16\pi t_0^{s}}
\frac{n\left[2(n-1)(s-2)-(D-2)(\omega+1)s^2\right]}{(s-2)}t^{s-2}
\end{equation}
and
\begin{equation}\label{P-BD}
p_{_{\rm BD}}=-\frac{\phi_0}{16\pi t_0^{s}}
\frac{n\left[2r(s-2)-(D-2)(\omega+1)s^2\right]}{(s-2)}t^{s-2},
\end{equation}
which, by applying~(\ref{s-om}), we have
\begin{equation}\label{ro}
\rho_{_{\rm BD}}=-\frac{\phi_0}{16\pi t_0^{^{[1-(D-1)r-n]}}}
\frac{(D-1)nr[(D-2)r-2n+2]+(D-4)n(n^2-1)}{[(D-1)r+n+1]t^{^{[(D-1)r+n+1]}}}
\end{equation}
and
\begin{equation}\label{P.i}
p_{_{\rm BD}}=\frac{\phi_0}{16\pi t_0^{^{[1-(D-1)r-n]}}}\frac{(D-1)(D-4)nr^2+n(n+1)
[(D-2)(n-1)-2r]}{[(D-1)r+n+1]t^{^{[(D-1)r+n+1]}}}.
\end{equation}
It is straightforward to show that the conservation law for the
above induced EMT is satisfied, as expected. Consequently,
from~relations~(\ref{ro}) and~(\ref{P.i}), the equation of state
for the power--law solutions on the $D$--dimensional hypersurface
is
\begin{eqnarray}\label{EOS.gen}
p_{_{\rm BD}}=W_{_{\rm BD}}\rho_{_{\rm BD}}, \hspace{10mm}
W_{_{\rm BD}}\equiv-\frac{(D-1)(D-4)r^2-(n+1)[2r-(D-2)(n-1)]}{(D-1)r[(D-2)r-2(n-1)]+(D-4)(n^2-1)}.
\end{eqnarray}

We proceed to discuss cosmological consequences
for different types of matter. Hence, it will be appropriate to express the
parameter $r$ in terms of the deceleration parameter
$q=-a\ddot{a}/(\dot{a})^2$, namely
 \begin{eqnarray}\label{rq}
r=\frac{1}{q+1}.
\end{eqnarray}
In order to proceed and analyze the induced quantities on the hypersurface, we
should express the exponent associated to the scalar $\psi$, the
scale factor of the $(D+1)$th dimension, in terms of $W_{_{\rm BD}}$, $r$ and
$D$. Thus, from relation~(\ref{EOS.gen}), we get
\begin{eqnarray}\label{n.gen}
 n=\frac{r[(D-1)W_{_{\rm BD}}+1]\pm \sqrt{\Delta}/2}{(D-4)W_{_{\rm BD}}+(D-2)},
\end{eqnarray}
where
\begin{eqnarray}\label{delta.gen}
 \Delta\equiv &\!\!\!+\!\!\!& 4r^2\Big[(D-1)W_{_{\rm BD}}+1\Big]-4\Big[(D-4)W_{_{\rm BD}}+(D-2)\Big]\\\nonumber
\times\Bigg\{&\!\!\!\Big[\!\!\!&(D^2-3D+2)W_{_{\rm BD}}+(D^2-5D+4)\Big]r^2
+2\Big[(D-1)W_{_{\rm BD}}-1\Big]r+\Big[(4-D)W_{_{\rm BD}}+2-D\Big]\Bigg\},
\end{eqnarray}
where $W_{_{\rm BD}}$ and $D$ must be accurately set, such that
we always have $\Delta\geq0$ and a non--vanishing value for the denominator of (\ref{n.gen}).
In addition, the resulted value for $n$ should give positive values for the induced energy density.

Consequently, let us re--write the reduced BD cosmological
power--law solution on a $D$--dimensional hypersurface as
\begin{eqnarray}\nonumber
ds^2&\equiv&dS^2{\Big|}_{_{\Sigma_{o}}}=
-dt^2+a_0^2\left(\frac{t}{t_0}\right)^{2r}\left[\sum^{D-1}_{i=1}
\left(dx^{i}\right)^{2}\right],\\\nonumber
\\\nonumber
\phi&=&\phi_0\left(\frac{t}{t_0}\right)^{[1-(D-1)r-n]},\\\nonumber
\\\nonumber
V(\phi)&=&\frac{n(D-2)\left[(D-1)r^2+n^2-1\right]}{\left[(D-1)r+n+1\right]
 t_0^2\phi_0^{^{2/[(D-1)r+n-1]}}}\phi^{[(D-1)r+n+1]/[(D-1)r+n-1]},\\\nonumber
 \\
\label{d-gen-solution}
\rho_{_{\rm BD}}&=&-\frac{\phi_0}{16\pi t_0^{^{[1-(D-1)r-n]}}}
\frac{(D-1)nr[(D-2)r-2n+2]+(D-4)n(n^2-1)}{[(D-1)r+n+1]t^{^{[(D-1)r+n+1]}}},
\end{eqnarray}
where $n$, as a function of the parameters $W_{_{\rm BD}}$, $r$ and $D$, is
given by~(\ref{n.gen}).  The constant $\phi_0$ has been assigned
to the value of the BD scalar field at some arbitrary fixed time
$t_0$.
%The above solution is indeed the $D$--dimensional extension
%of four--dimensional solution obtained in~\cite{Ponce1}.

In order to study accelerating solutions (in particular for late times)
we further discuss the energy density and the pressure associated to
the BD scalar field.
From~(\ref{BD-Eq-DD}) and~(\ref{ohanlon metric}),
the FLRW equations on a $D$--dimensional hypersurface
can be written as\footnote{We should note that the
energy density and pressure associated to the
BD scalar field, according to some conventional notations,
are denoted by $\rho_{\phi}$ and $p_{\phi}$, respectively; and they are not derived from the
second component of the induced EMT, i.e. $T_{\mu\nu}^{^{[\rm \phi]}}$.}
\begin{eqnarray}
\label{DD-FRW-eq1}
\frac{(D-1)(D-2)}{2}H^2
\!\!&=&\!\!\frac{1}{\phi}\left(8\pi\rho_{_{\rm BD}}+\rho_{\phi}\right),\\
\label{DD-FRW-eq2}
(D-2)\frac{\ddot{a}}{a}+\frac{(D-2)(D-3)}{2}H^2
\!\!&=&\!\!-\frac{1}{\phi}\left(8\pi p_{_{\rm BD}}+p_{\phi}\right),
\end{eqnarray}
where $H=\dot{a}/a$ is the Hubble expansion
rate of the universe, $\rho_{_{\rm BD}}$ and $p_{_{\rm BD}}$ are given by
(\ref{ro}) and (\ref{P.i}), respectively, and the energy density
and pressure of the BD scalar field~\cite{T02} are given by
\begin{eqnarray}
\label{rho-phi-gen}
\rho_\phi\!\!&\equiv\!\!&\frac{\omega}{2}
\frac{\dot{\phi}^2}{\phi}+\frac{V(\phi)}{2}-(D-1)H\dot{\phi},
\label{p-phi-gen}\\
p_\phi\!\!&\equiv\!\!&\frac{\omega}{2}
\frac{\dot{\phi}^2}{\phi}-\frac{V(\phi)}{2}+\ddot{\phi}+(D-2)H\dot{\phi}.
\end{eqnarray}
Employing~(\ref{power-law-phi}), we get
\begin{equation}\label{ro-phi}
\rho_{\phi}=\frac{\phi_0}{t_0^{s}}
\frac{s\left[\omega s(s-2)+(D-2)(\omega+1)ns-2(D-1)(s-2)r\right]}{2(s-2)}t^{s-2}
\end{equation}
and
\begin{equation}\label{p-phi}
p_{\phi}=\frac{\phi_0}{t_0^{s}}
\frac{s\left[\omega s(s-2)-(D-2)(\omega+1)ns+2(D-2)(s-2)r+2(s-1)(s-2)\right]}{2(s-2)}t^{s-2},
\end{equation}
in which the equation of state for the BD scalar field is described by
\begin{eqnarray}
\label{w-phi}
p_\phi=W_\phi\rho_\phi \hspace{5mm}{\rm where}\hspace{5mm}
W_{\phi}\equiv\frac{\omega s(s-2)-(D-2)(\omega+1)ns+2(D-2)(s-2)r+2(s-1)(s-2)}
{\omega s(s-2)+(D-2)(\omega+1)ns-2(D-1)(s-2)r}.
\end{eqnarray}

Equation~(\ref{DD-FRW-eq1}) can be written in the form $\Omega_{_{\rm BD}}+\Omega_{\phi}=1$,
with
\begin{eqnarray}
\label{density.par.def}
\Omega_{_{\rm BD}}&\equiv&\frac{16\pi}{(D-1)(D-2)}\frac{\rho_{_{\rm BD}}}{H^2\phi},\\\nonumber
\Omega_{\phi}&\equiv&\frac{2}{(D-1)(D-2)}\frac{\rho_{\phi}}{H^2\phi},
\end{eqnarray}
where $\Omega_{_{\rm BD}}$ and $\Omega_{\phi}$ are the density parameters associated to
the reduced matter and the BD scalar
field\footnote{Such definitions have been applied in~\cite{DP99}}, respectively.
By employing~(\ref{power-law-phi}), (\ref{ro-BD}), (\ref{ro-phi}),
the density parameters are written as
\begin{eqnarray}
\label{density.par}
\Omega_{_{\rm BD}}&=&\frac{n\left[2(n-1)(s-2)-(D-2)(\omega+1)s^2\right]}{(D-1)(D-2)(s-2)r^2},\\\nonumber
\Omega_{\phi}&=&\frac{s\left[\omega s(s-2)+(D-2)ns(\omega+1)-2r(D-1)(s-2)\right]}{(D-1)(D-2)(s-2)r^2}.
\end{eqnarray}

As the main purpose of the IMT framework is
to show that the matter in the universe
has a geometrical origin~\cite{stm99}; in the following subsections,
we will examine the $D$--dimensional solutions for some values of $W_{_{\rm BD}}$.
In order to compare the resulted solutions
with the ones obtained in the conventional BD theory
(with or without any scalar potential or other
generalization, e.g. a variable BD coupling parameter), as well as with
observational data, somewhere, we will restrict ourselves to $D=4$.
%{\bl We should note that all the following allowed ranges for
%the BD coupling parameter, according to respect

\subsection{Modified Brans--Dicke Cosmology and Quintessence}

The observational data have shown that
we are living in an accelerated expanding universe.
%However, the Friedmann-Einstein acceleration equation
%gives an accelerated expansion for the universe
%if and only if the pressure of is negative.
%Such a kind of matter with this exotic property has been called dark energy.
A lot of cosmological quintessence models of dark
energy~(see e.g.~\cite{PB99,T02,Faraoni.book,P13} and references therein),
have been presented to explain the present epoch of the universe. Scalar-tensor
models of dark energy have been known as extended quintessence~\cite{FJ06}.
As the most of the mentioned models based on the phenomenological basis,
thus, many endeavors also have been done to extract this property for the
universe from fundamental physics. For instance, in several investigations, when the
conventional BD theory has been used in quintessential scenarios,
 a scalar potential has been included by hand.
 As we assert that the MBDT, and in turn the induced scalar potential,
is constructed based on fundamental principles,
one of the aims of the MBDT is to present a model
to explain the accelerated expansion.
In the following subsections, we would like to examine
the MBDT cosmology for describing the various epoches of the universe.

\begin{figure}
\centering{}\includegraphics[width=4in]{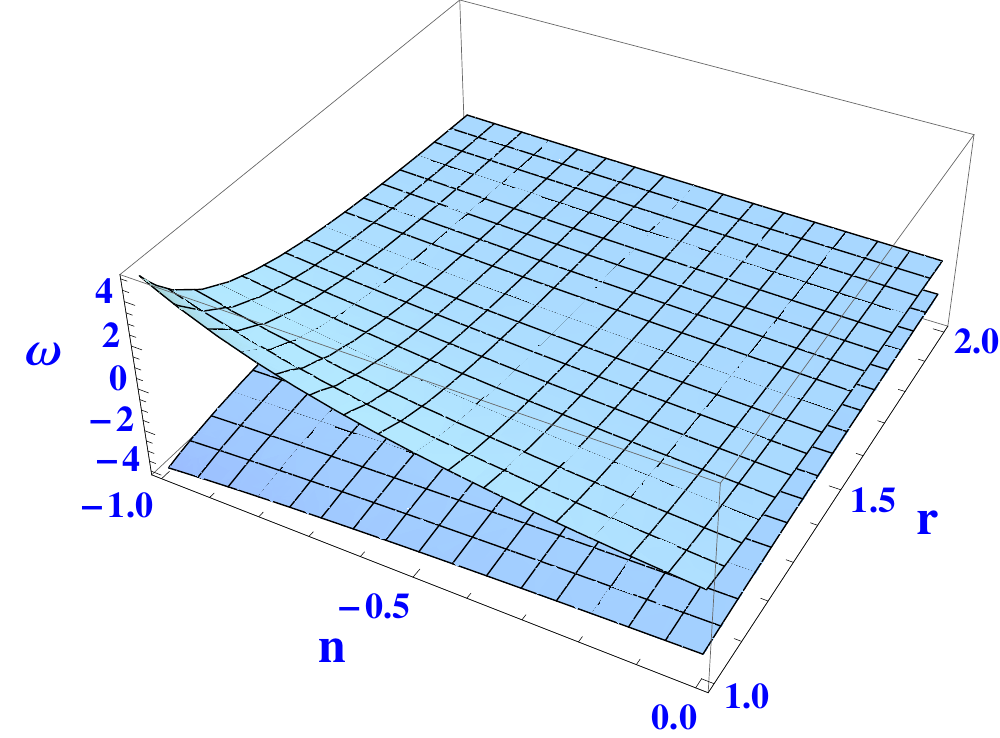}\hspace{4mm}
\caption{{\footnotesize The region between two hypersurfaces shows the allowed region
for $\omega$ versus $n$ and $r$ (the permitted values) for an accelerated expansion.}}
\foreignlanguage{english}{\label{quint.omega-n-r.fig}}
\end{figure}

For an accelerating universe, from the power--law
solution~(\ref{power-law-phi}), we must have $r>1$.
 Besides,
 we assume\footnote{In the subsections~\ref{MDU},~\ref{RDU} and~\ref{Nariai}, we will assume
 different choices of $W_{_{\rm BD}}$.} the equation of the state parameter of the baroscopic matter
 is restricted to be between $0$ and $1$.
 Thus, from~(\ref{EOS.gen}) with $D=4$, we get
$-1<n<r-1$ or equivalently, from~(\ref{s-om}), we have $2-4r<s<2-3r$,
which implies that $s$ is always negative.
As we are interested in a shrinking fifth dimension, we would like to restrict
the solutions to only negative values of $n$, namely, we take $-1<n<0$, which gives $1-3r<s<2-3r$.
Furthermore, in order to satisfy the positivity conditions for $\rho_{_{\rm BD}}$
and $\rho_\phi$, from relations~(\ref{ro-BD}) and~(\ref{ro-phi}),
and the resulted negative values for $s$ and $n$,
the BD coupling parameter must be restricted as
\begin{eqnarray}\label{omega-s.quint-2}
\frac{2(1-s)}{s}<\omega<\left[\frac{(3r+s)(2-s)}{s^2}-1\right],
 \hspace{5mm} {\rm with} \hspace{5mm} 1-3r<s<2-3r,
\end{eqnarray}
or equivalently, from relation~(\ref{s-om}), we get
\begin{eqnarray}\label{omega-n.quint-2}
\frac{2(3r+n)}{1-3r-n}<\omega<\left[\frac{(1-n)(1+3r+n)}{(1-3r-n)^2}-1\right],
\hspace{5mm}{\rm with} \hspace{5mm} -1<n<0.
\end{eqnarray}
According to the above relations, in Fig.~\ref{quint.omega-n-r.fig},
the allowed region for the BD coupling parameter,
(the region between the two hypersurfaces) has been specified versus $n$ and $r$ for an accelerated
expansion.
%Also, in Fig.~\ref{n-om1.fig}, we have plotted the allowed ranges of the
%BD coupling parameter in $(n,\omega)$ parameter space
%for $r=1.05$ and $r=1.15$.
This figure has been plotted by respecting
\footnote{Also,
in~\cite{SSS01}, in the context of power law cosmology,
by comparing the results of the scenario with the ultra--compact radio and also with SNIa, the best fit
values for the different parameters have been obtained. It was shown that, in the mentioned model,
the best fit value for $r$ is approximately~$1.25$.}
$r>1$ and the weak energy condition of
the matter associated to the BD scalar field and the induced matter.
%\begin{figure}
%\centering{}\includegraphics[width=3.3in]{nomegaQ1}\hspace{4mm}
%\includegraphics[width=3.3in]{nomegaQ2}\hspace{4mm}
%\caption{{\footnotesize The allowed ranges
%for $\omega$ in the $(n,\omega)$ parameter
%space for $r=1.05$ (the left figure) and $r=1.15$ (the right figure),
 %in which the weak energy conditions are satisfied.}}
%\foreignlanguage{english}{\label{n-om1.fig}}
%\end{figure}
%\bl{By employing the relations~(\ref{}), one can easily show that $W_{\phi}$
%can be written versus $n$, $r$ and density parameters as
%\begin{eqnarray}\label{Wphi-other}
%W_\phi=-\frac{1}{8\pi}\left[\frac{9r^3+3(n-1)r^2+(n+1)(n+2)r+n^3-n}
%{6rn(n-r-1)}\right]\frac{\Omega_{_{\rm BD}}}{\Omega_{\phi}}.
%\end{eqnarray}
In the following figures, we will find a few allowed ranges for
the parameters of the model, which not only do they respect to the mentioned
physical conditions, but also match with the recent observational data for the BD theory~\cite{LWC13}.

\begin{figure}
\centering{}\includegraphics[width=4in]{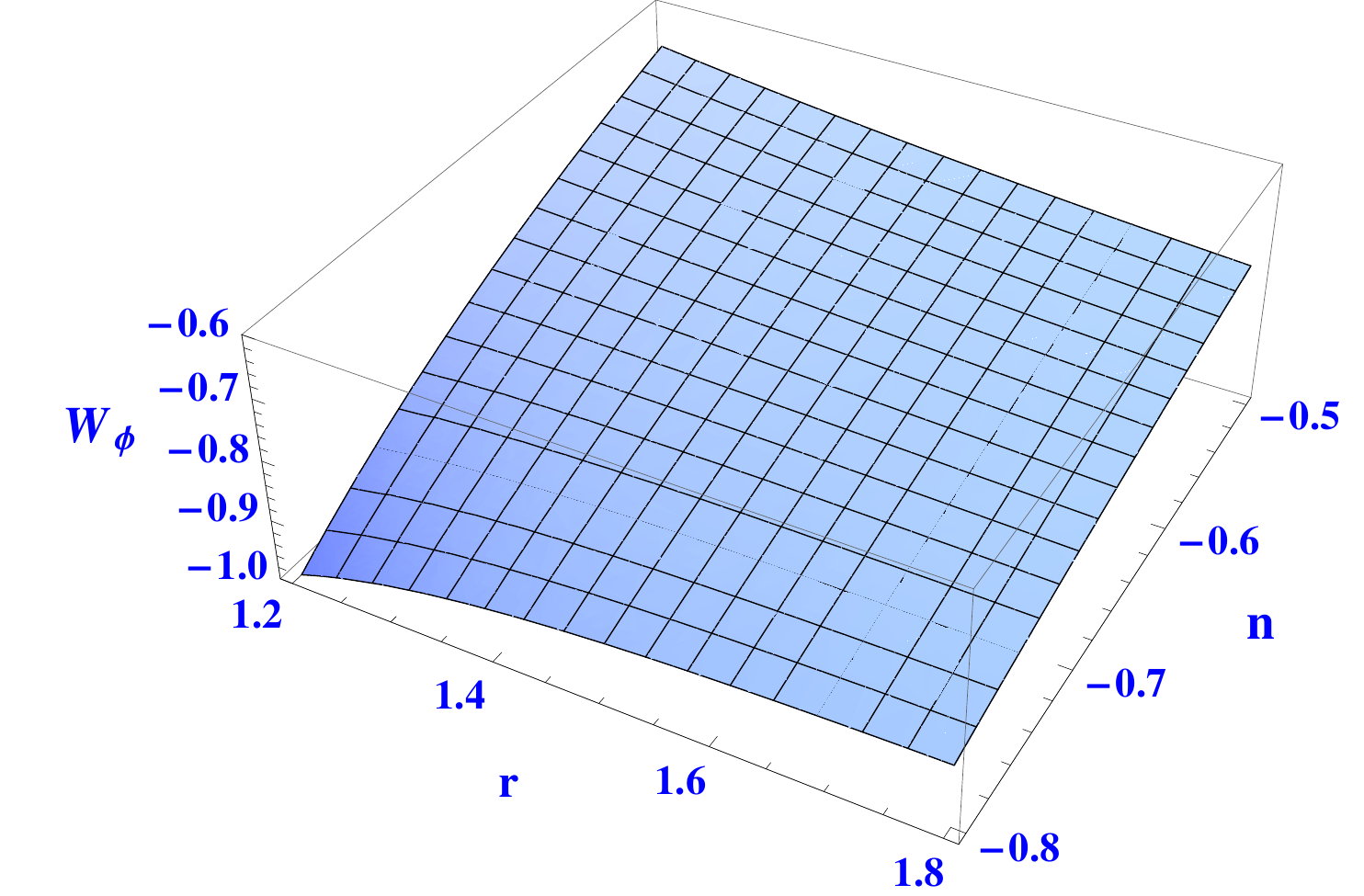}\hspace{4mm}
\caption{{\footnotesize Resulted ranges
for $W_\phi$ versus $r$ and $n$ in their corresponding allowed ranges.}}
\foreignlanguage{english}{\label{wphi-n-r.fig}}
\end{figure}

By employing Eqs.~(\ref{s-om}) with $D=4$, we have
obtained $W_\phi$ versus $n$ and $r$, and then specified
an allowed range for it, numerically.
As a sample, Fig.~\ref{wphi-n-r.fig} gives an acceptable ranges for $W_\phi$ which may in
agreement with the recent observational data
describing the current
accelerated universe. Especially, this figure also indicates that the obtained ranges for
$W_\phi$ as well as $r$ are in accordance with
the results of conventional BD theory~\cite{SS01, SS03}.

\begin{figure}
\centering{}\includegraphics[width=4in]{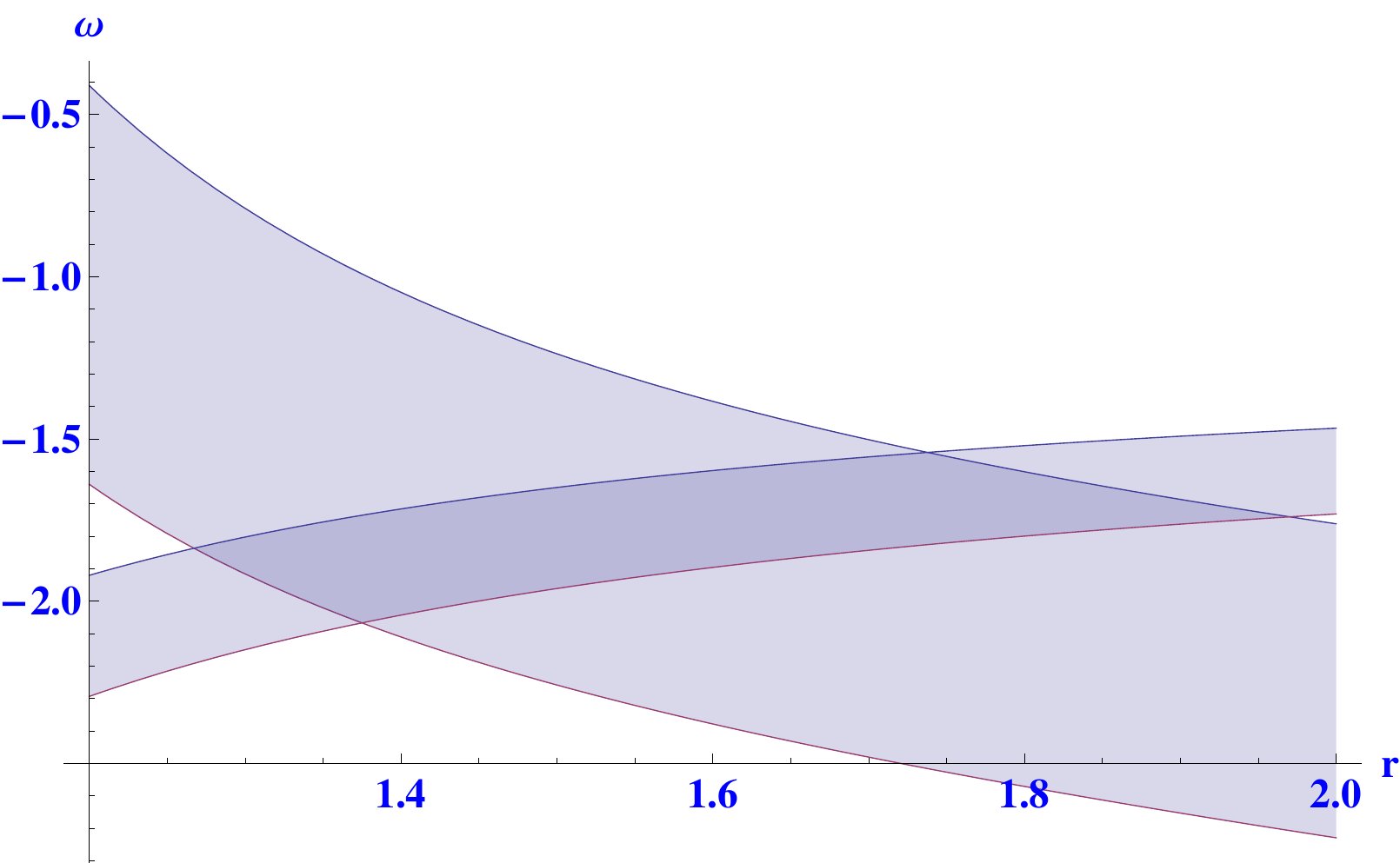}\hspace{4mm}
\caption{{\footnotesize The overlap of the two spaces shows
the allowed region in the $(r,\omega)$ parameter space for $n=-0.9$,
$0.3<\Omega_{_{\rm BD}}<0.5$ and $-0.6<W_{\phi}<-0.8$.}}
\foreignlanguage{english}{\label{quint-obser-omega-r-n.fig}}
\end{figure}

In the other numerical endeavors, we have tried to find
allowed ranges for the BD coupling parameter. In this manner, we have used
the recent data for the present parameters
in the model\rlap.\footnote{The best fit values of density
parameters associated to the dark energy and matter
  are $0.6952$ and $0.3048$, respectively~\cite{LWC13}.}
 The results of our study have shown that the best range for
 the BD coupling parameter is almost constrained as $-3/2<\omega<-2$.
 For instance, in Fig.~\ref{quint-obser-omega-r-n.fig}, we have plotted $\omega$ versus $r$
 for an allowed value of $n$.
 This result is also in accordance with the consequences obtained
 by the mentioned conventional BD theory.

\begin{figure}
\centering{}\includegraphics[width=3.3in]{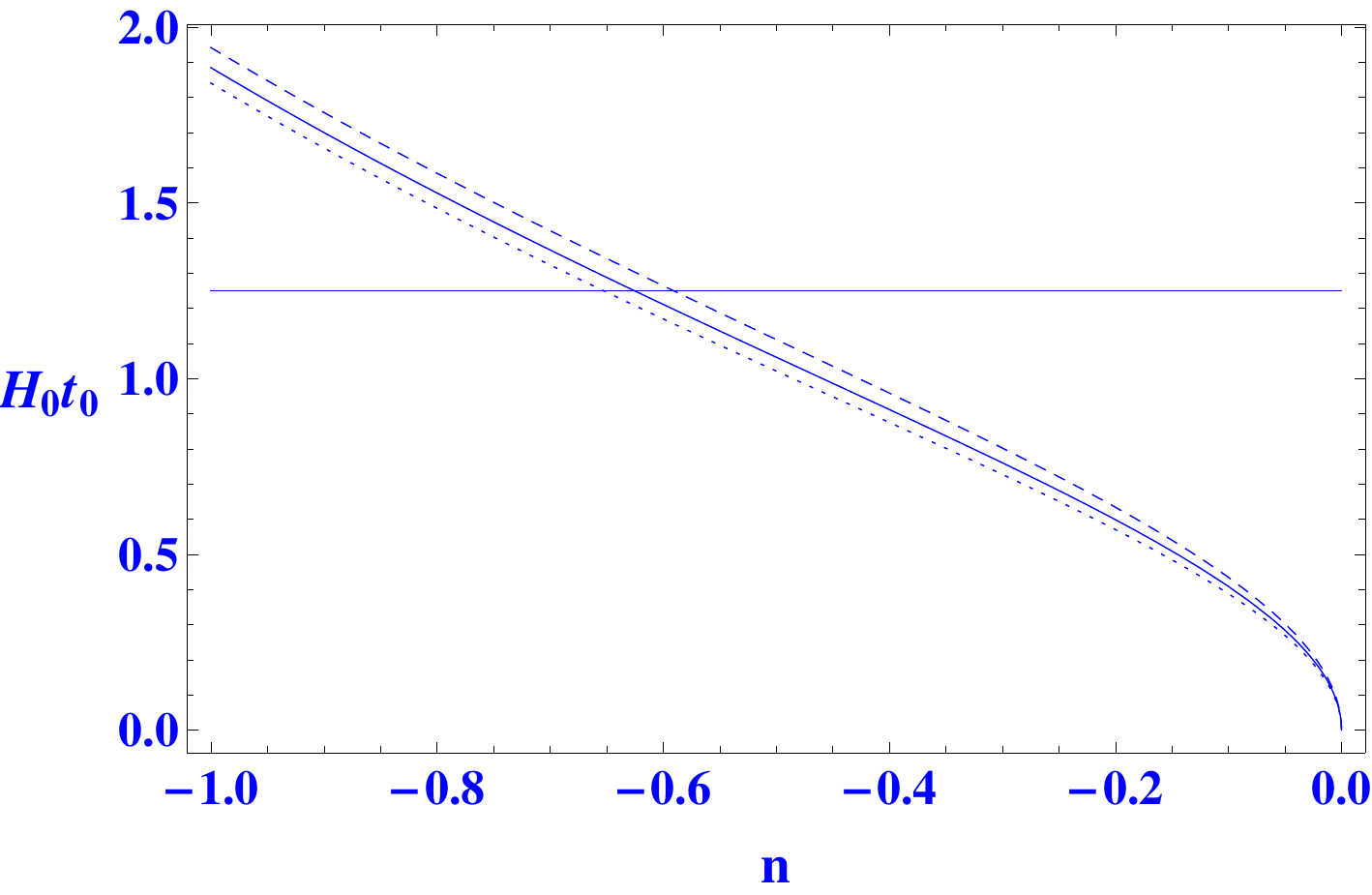}\hspace{4mm}
\includegraphics[width=3.3in]{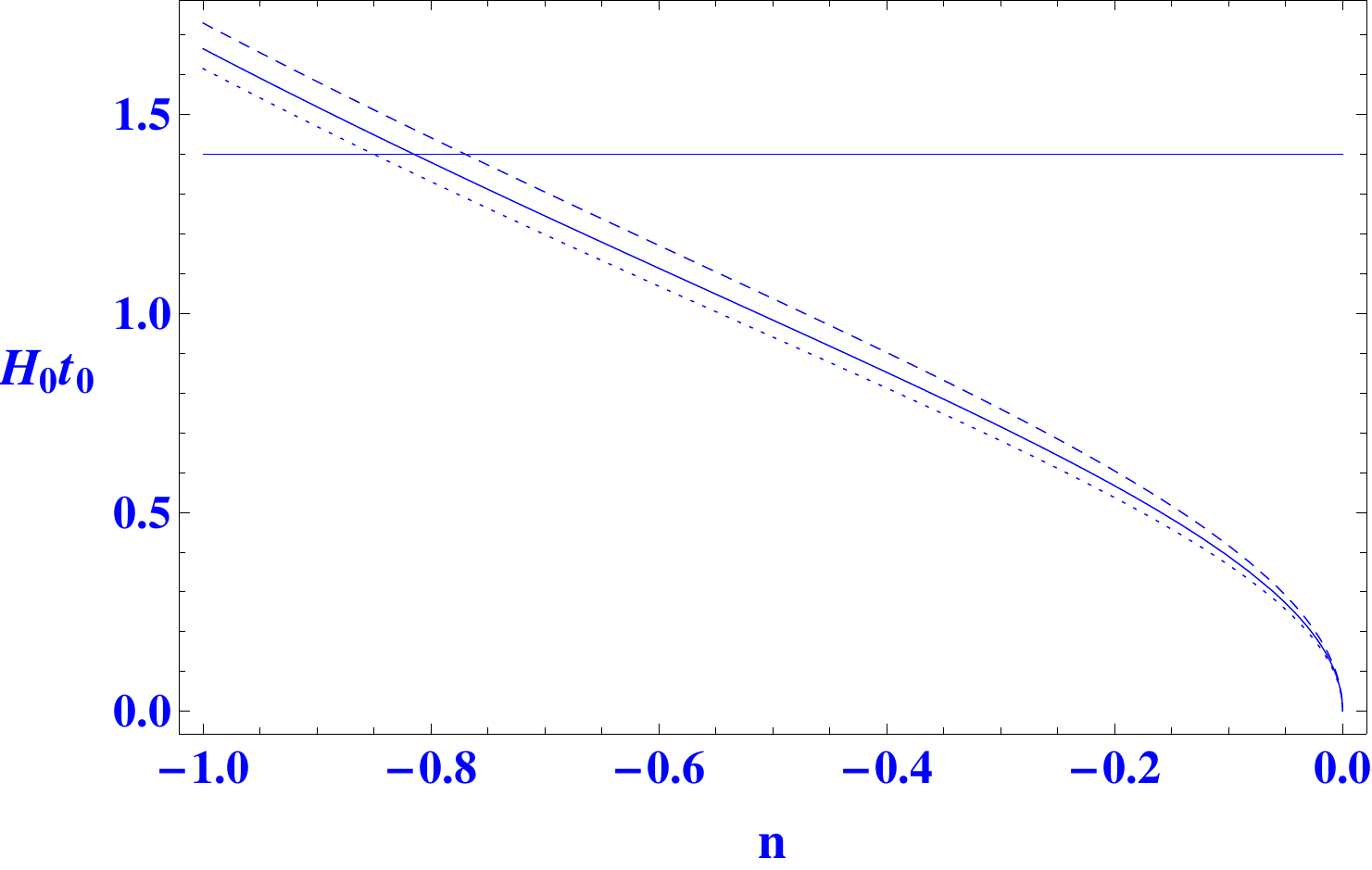}\hspace{4mm}
\caption{{\footnotesize $H_0t_0$ versus $n$ with $r=1.25$ (left figure) and $r=1.40$ (right figure)
for $\omega=-2$ (the dashed curve), $\omega=-1.8$ (the solid curve)
and $\omega=-1.65$~(the dotted curve). We set $\Omega_{_{\rm BD}}=0.3048$.}}
\foreignlanguage{english}{\label{Age.fig}}
\end{figure}

%In addition to the above restrictions on the
%BD coupling parameter and $n$, we want to
%find the allowed regions for the parameters of the model \bl{such} that
%\bl{they will be} compatible with recent observational data for an accelerated universe.
%We remind that the parameter $r$, for the mentioned
%reasons, should be very close to unity, and
%again the mentioned physical conditions should be considered.
%Thus, by using equations~(\ref{w-phi}) and~(\ref{density.par}),
%we have plotted
 %Fig.~\ref{Q-omega-n.fig}. Therein, we have plotted
%the allowed ranges (the overlap region) in $(n, \omega)$
%parameter space for particular limits of
 %the density parameter (associated to the induced matter) as well as
%equation of state parameter for the BD scalar field.
%\begin{figure}
%\centering{}\includegraphics[width=3.3in]{quin1}\hspace{4mm}
%\includegraphics[width=3.3in]{quin2}\hspace{4mm}
%\caption{{\footnotesize $\omega$ versus $n$
%for $r=1.05$ (the left figure) and $r=1.15$ (the right figure),
 %in which $\Omega_{_{\rm\phi}}$ and $\Omega_{_{\rm BD}}$ are restricted
 %as $-0.8<\Omega_{_{\rm\phi}}<-0.6$ and $0.3<\Omega_{_{\rm BD}}<0.5$.}}
%\foreignlanguage{english}{\label{Q-omega-n.fig}}
%\end{figure}
Since in scalar--tensor theories, the gravitational constant is varying with time,
the gravitational coupling (in the Jordan frame) is constrained by the gravitational experiments.
From the effective BD action on the hypersurface, the Newton's
gravitational constant is read as the inverse of the BD scalar field, $G_{\rm N}=\phi^{-1}$.
However, $G_{\rm N}$ cannot be interpreted, physically, the same as Newton's gravitational constant in GR.
In fact, it is included in the Newton force (as determined by Cavendish--type experiments)
 as $G_{\rm eff}m_1m_2/r_{12}^2$ in which $m_1$ and $m_2$
are two close test masses and $r_{12}$ is their distance. Hence,
the effective gravitational constant in the BD theory (in the weak field limit) is
given by $G_{\rm eff}=\phi^{-1}(2\omega+4)/(2\omega+3)$~\cite{EP01,NP07}.
Thus, the rate of variation of $G_{\rm N}$ and $G_{\rm eff}$ have the same value and, the
recent experimental limits~\cite{LWC13} have shown that, at present time, it reads
\begin{eqnarray}\label{G-limit}
\left[\frac{\dot{G}_{\rm eff}}{G_{\rm eff}}\right]_{t=t_0}\approx1.42\times10^{-13}yr^{-1},
\end{eqnarray}
where $t_0$ is the age of the universe.
In our model, we have
\begin{eqnarray}\label{G-limit-mbdt}
\left[\frac{\dot{G}_{\rm eff}}{G_{\rm eff}}\right]_{t=t_0}=-\frac{s}{t_0}=-\frac{s}{r}H_0,
\end{eqnarray}
where $H_0=r/t_0$ is the Hubble parameter at present.
Therefore, the parameters $s$ and
$r$, for an accelerated expansion at late times,
should satisfy the constraint~(\ref{G-limit}).
If we assume that the age of the universe to be $13.7179$ Gyrs (as a best fit value),
as estimated in~\cite{LWC13}, thus in our model, $H_0t_0=r$ may approach to one.
On the other hand, for a four--dimensional space--time,
from (\ref{power-law-phi}), (\ref{ro-BD}) and (\ref{density.par.def}), we have
\begin{eqnarray}\label{H0t0}
(H_0t_0)^2=\frac{n\left[(n-1)(1+3r+n)+(\omega+1)(1-3r-n)^2\right]}{3(1+3r+n)\Omega_{_{\rm BD}}}.
\end{eqnarray}
In Fig.~\ref{Age.fig}, $H_0t_0$ has been plotted versus $n$ (in the allowed range)
for some allowed values of $\omega$ and\footnote{Further to the
mentioned argument for an approximate value of $r$,
other investigations~\cite{KSTW99,SBL99},
which are also based on the power law cosmological approaches,
have shown that the parameter $r$ is of order unity.}
 $r$. As it is seen, in the permissable ranges of the parameters of the model,
all the curves associated to $H_0t_0$ intersect at least once the line associated to
constant values of $r=1.25$ and also $r=1.40$.

\subsection{Matter--Dominated Universe}
\label{MDU}
For a matter--dominated universe, we set $W_{_{\rm BD}}=0$, thus
relations~(\ref{n.gen}) and (\ref{delta.gen}) reduce to
\begin{eqnarray}\label{n.gen.matt}
n=\frac{r\pm\sqrt{\Delta_{\rm matt}}}{(D-2)} \hspace{5mm} {\rm with} \hspace{5mm}
\Delta_{\rm matt}\equiv-\left[D(D^2-7D+14)-9\right]r^2+2(D-2)r+(D-2)^2.
\end{eqnarray}
Therefore, by substituting these values of $n$ into
solution~(\ref{d-gen-solution}), we obtain the solutions associated with a
matter--dominated universe, in which all of the induced quantities
are in terms of the parameters $D$ and $r$. Let us investigate the
solutions for a four--dimensional space--time. By substituting
$D=4$ into relation~(\ref{n.gen.matt}), we get
\begin{eqnarray}\label{n.gen.matt.4d}
n=\frac{1}{2}(r\pm|r+2|),
\end{eqnarray}
where it yields $n=-1, r+1$.
As the case where $n=r+1$ gives $\rho=0$ (for $D=4$), it is not acceptable.
In what follows, we derive the
solution for the case where $n=-1$.

 By substituting $n=-1$ and $D=4$ into
 relations~(\ref{s-om}) and (\ref{d-gen-solution}) we obtain
\begin{eqnarray}\label{s-om-matt-4d}
s&=&2-3r,\hspace{27mm} \omega=-\frac{4(3r^2-3r+1)}{(2-3r)^2}\\\nonumber
\rho&=&\frac{\phi_0 (r+2)}{8\pi t_0^{2}}\left(\frac{t}{t_0}\right)^{-3r}, \hspace{7mm}
V=-\frac{2\phi_0r}{t_0^2}\left(\frac{t}{t_0}\right)^{-3r}.
\end{eqnarray}
In a particular case where $\mid\omega\mid$ tends to infinity,
we obtain $r=2/3$ and thus $s=0$. Namely, in this case, the BD
scalar field takes constant values and, without loss of
generality, from~(\ref{d-gen-solution}), we can assume $V=0$,
which implies that the usual spatially flat FLRW cosmology
(for a matter--dominated universe) in the
context of GR has been recovered. Besides, $n=-1$ indicates that
the extra dimension contracts as the cosmic time increases.

Let us proceed to probe the general solutions without assuming any
particular value for $\omega$. From relations~(\ref{rq})
and~(\ref{d-gen-solution}), the decelerated parameter can be
written in terms of $\omega$ as
\begin{equation}\label{q-matt-4d}
q=\frac{1+\omega\pm\sqrt{-3(1+\omega)}}{2(1+\omega)},
\end{equation}
which implies that, in order to get real values for $q$ (or $r$),
the BD coupling parameter must be restricted to $\omega<-1$.
Also, expressing the above relation in terms of $r$ leads to the other restriction on
the BD coupling parameter, namely, $\omega\neq-4/3$.
%The former range does~not yield any restriction on
%$q$, but the latter implies that $q$ must be restricted to
%either\footnote{We emphasize that some of these results herein the $n=-1$
%case, have been also obtained in~\cite{Ponce1}. However, in
%order to have complete set of solutions, we have mentioned these
%results with additional physical instructions from
%the herein $D$--dimensional perspective of a MBDT setting.}\
 %$q<(1-\sqrt{6})/2$ or $q>(1+\sqrt{6})$.
On the other hand, from relations~(\ref{ro-BD}) and (\ref{ro-phi}), the positivity condition
for $\rho_{_{\rm BD}}$ and $\rho_{\phi}$, in the case where $n=-1$ and $r>1$, are given by
\begin{equation}\label{WE-matt}
-\frac{2(3r-1)}{3r-2}<\omega<\frac{6r}{(2-3r)^2}-1.
\end{equation}

\begin{figure}
\centering{}\includegraphics[width=4in]{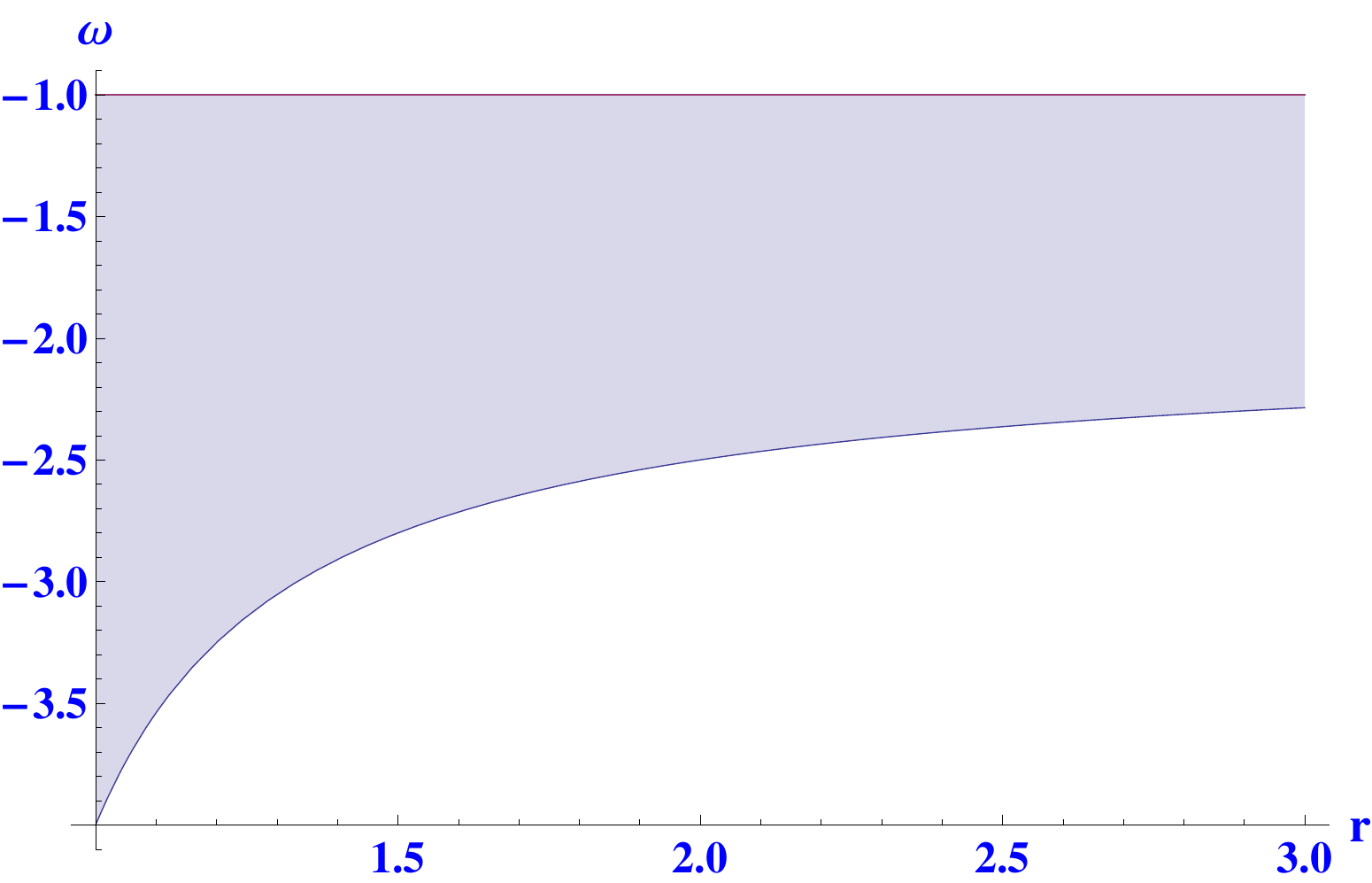}\hspace{4mm}
\caption{{\footnotesize The allowed range of the BD coupling parameter in $(r,\omega)$ parameter space
for $n=-1$ and small values of $r$ for a matter--dominated universe.}}
\foreignlanguage{english}{\label{we-matt-fig}}
\end{figure}

Thus, according to the above physical conditions\rlap,\footnote{Unlike~\cite{Ponce1},
we do not assume $\omega$ to be restricted to
$\omega>-3/2$.}
we have plotted this range of the BD coupling
parameter versus $r$ in Fig.~\ref{we-matt-fig}.
We should mention that this specified range of $\omega$ satisfies
the weak energy conditions for the
matter associated to the BD scalar field and the induced matter.
In addition, we notice that the obtained range of
the $\omega$ is applicable to produce values for $q$ which are in agreement with the current
observed measurements for the present epoch of the universe, namely, $q=-0.67\pm0.25$.
%Whereas the latter range of $q$ leads us to a
%decelerated universe.
As $n=-1$, note that while the usual spatial
dimensions expand with $t$, the extra dimension contracts.

\begin{figure}
\centering{}\includegraphics[width=4in]{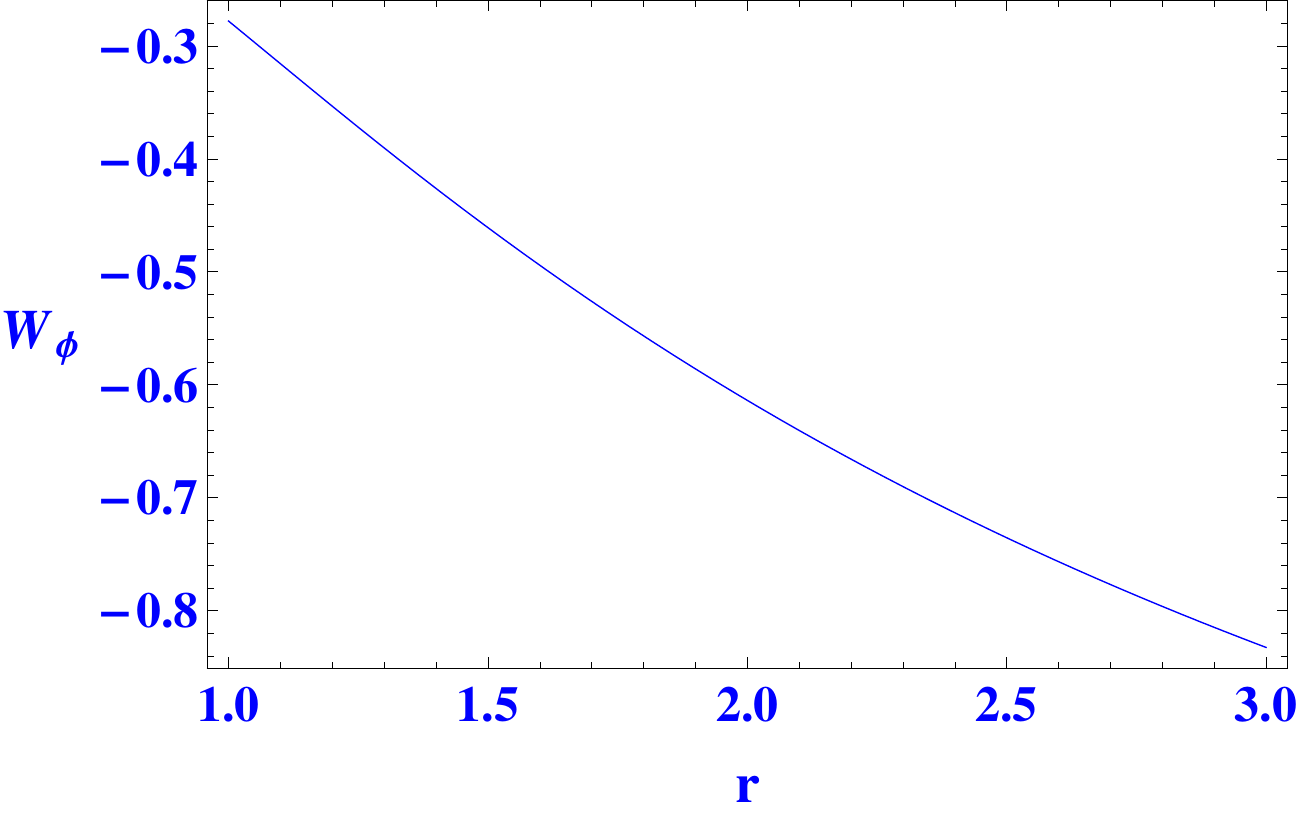}\hspace{4mm}
\caption{{\footnotesize $W_\phi$ versus $r$ for $\omega=-1.49$ for the matter dominated epoch.}}
\foreignlanguage{english}{\label{wphi-r-matt}}
\end{figure}

In Fig.~\ref{wphi-r-matt}, we have shown $W_{\phi}$ versus $r$ for a particular
value of the BD coupling parameter in the permissable range. This figure indicates that
if any permissable value of $\omega$ is taken, then for small values of
$r$, the value of $W_\phi$ can be consistent with the recent observational data.

 %{\bf $n=r+1$:} By applying the above approach for this value of $n$, we obtain
%\begin{equation}\label{s-om-matt-4d-2}
%s=-4r \hspace{10mm} {\rm and} \hspace{10mm}
%\omega=-\frac{1}{4}\left(5+\frac{1}{2r}\right),
%\end{equation}
%where $\omega\neq-5/4$.
%When $\mid\!\omega\!\mid$ goes to infinity, $r$ and $s$ vanish,
%namely, for large values of the BD coupling parameter, $\phi$
%takes a constant value and we have a static universe. However, in
%general, the deceleration parameter can be written in terms of
%$\omega$ as
%\begin{equation}\label{q-matt-4d-2}
%q=-(8\omega+11),
%\end{equation}
%where for the values of the BD coupling parameter restricted to
% $\omega>-11/8$, we have $q<0$.
%On the other hand, for $\omega>-(1+10r)/8r$
%the weak energy conditions are satisfied. In summary, in the case where $n=r+1$
% for $\omega>-11/8$ (in which $\rho_{_{\rm BD}}>0$ and $\rho_{\phi}>0$),
%we obtain another accelerated expansion.
%owever, we note that the extra dimension increases with $t$, the cosmic time.

In order to show that the cosmology based on MBDT can
work, appropriately, as an unified model for
describing dark matter--dark energy, besides the
above resulted solutions for only $r>1$, let
us apply the following approach
for explaining the consequences associated to the other values of $r$.
% for the case where the induced pressure is zero.}

\begin{figure}
\centering{}\includegraphics[width=4in]{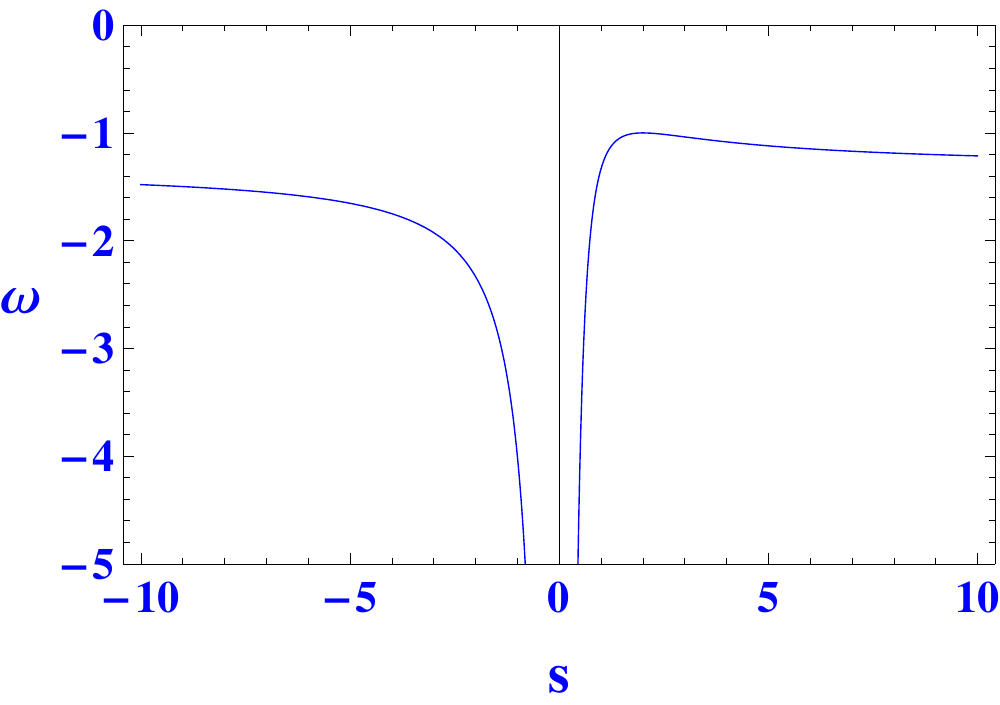}\hspace{4mm}
\caption{{\footnotesize $\omega$ versus $s$ for $n=-1$.
The right branch corresponds to a decelerating
expansion and the left one corresponds to an accelerating expansion.
When $s$ goes to zero, then the corresponding IMT solution is recovered.}}
\foreignlanguage{english}{\label{omega-s-DM-DE}}
\end{figure}

From equations~(\ref{DD-FRW-eq1}) and~(\ref{DD-FRW-eq2}), we get
\begin{eqnarray}
\label{DD-FRW-shetab}
\frac{\ddot{a}}{a}
=-\frac{1}{(D-1)(D-2)\phi}\Big[(D-3)\rho_{_{\rm tot}}+(D-1)p_{_{\rm tot}}\Big],
\end{eqnarray}
where $\rho_{_{\rm tot}}=8\pi\rho_{_{\rm BD}}
+\rho_{\phi}$ and $p_{_{\rm tot}}=8\pi p_{_{\rm BD}}+p_{\phi}$.
 As we have assumed $\phi>0$,
%thus, the terms inside the brackets, i.e.,  represents
%the strong energy condition. More precisely,
the sign of $(D-3)\rho_{_{\rm tot}}+(D-1)p_{_{\rm tot}}$ determines to have
whether accelerating or decelerating expansion (contraction).
 Let us focus on the the specific case of this subsection.
 Therefore, by using the relations~(\ref{s-om}), (\ref{ro-BD}),
 (\ref{P-BD}), (\ref{ro-phi}), (\ref{p-phi}) and setting
$D=4$ and $n=-1$ in
equation (\ref{DD-FRW-shetab}), we easily find that
 $\ddot{a}/a<0$ for $-1<s<2$ (except $s=0$) and $\ddot{a}/a>0$ for $s<-1$.
 Namely, the former case gives a decelerating expansion while the latter
 one gives an accelerating expansion for the universe.
 Note that when $s>2$, we have also $\ddot{a}/a>0$,
 but in this case, we have an accelerating contraction. Moreover, we
 have $\ddot{a}/a=0$ (zero acceleration) when $s=-1$ which
 corresponds to $r=1$ and $\omega=-4$. For the case where $s=2$, we get $r=0$ that yields a static universe.
 %When $s=2$, we get $r=0$ and $\omega=-1$, respectively.
 For both the ranges $-1<s<2$ and $s<-1$, we get $\rho_{_{\rm tot}}>0$ whilst,
 as Figure~\ref{omega-s-DM-DE} shows, $\omega$ takes small negative values such that
 the allowed ranges (of $\omega$) are almost the same for both of the branches.
 Note that in this figure, when $s$ tend to zero
 then $|\omega|$ takes very large values.
 Hence, $\phi$ takes constant values and, consequently, the
 corresponding solutions in the IMT is recovered.
  Although the observational
 viability of the case $s<-1$ (which corresponds to $r>1$) has already been discussed,
 but, once again we note that, when $s$ is restricted to the range $s<-1$, then the deceleration
 parameter is restricted to $-1<q<0$, while for $-1<s<2$, it takes positive values.
 We should note that, in the phenomenological models based on the
 BD theory,
 %corresponding with the different phases that
 %should be described (by a pressureless matter),
 the scalar potential is imposed
 by hand. Instead, in our model, the induced
 scalar potential
 %corresponding to each phase,
 is automatically procreated by the geometry of the extra dimension.
 and is responsible for describing each epoch.
 %which the resulted solutions, fortunately, are in
 %accordance with the recent observational data.}

\subsection{Radiation--Dominated Universe}
\label{RDU}
For the radiation--dominated case in a $D$--dimensional
hypersurface, we set $W_{_{\rm BD}}=1/(D-1)$ in~(\ref{n.gen}) which gives
\begin{eqnarray}\label{n.gen.rad}
n=\frac{2(D-1)r\pm\sqrt{\Delta_{\rm rad}}}{D(D-2)-2} \hspace{5mm}
{\rm with} \hspace{5mm} \Delta_{\rm rad}\equiv
D^2(D-4)\left[D-(D-2)(D-1)r^2\right]+8D+4.
\end{eqnarray}

In the special case where $D=4$, relation~(\ref{n.gen.rad}) gives
$n=r\pm1$. As mentioned, the case $n=r+1$ is not acceptable.
In the following, we explain the solutions associated to $n=r-1$.
%i) $n=r+1$: As for a decelerated expansion, $r$ is restricted to
%$0<r<1$. For satisfying $\rho_{_{\rm BD}}>0$ and $\rho_{\phi}>0$, we again obtain
%$\omega>-(1+10r)/8r$. On the other hand, in order to have a decelerated universe, we get
% $\omega<-11/8$. Thus for this case, if $\omega$ to be
% restricted to $-(1+10r)/8r<\omega<-11/8$, the solution is applicable for
%radiation dominated universe. We should mention that for this
%case, the same as for accelerated expansion, the fifth dimension does not contract with time.
%In Fig.~\ref{WE-rad1}, the allowable range of $\omega$ for $0.1<r<1$ has been plotted.
%\begin{figure}
%\centering{}\includegraphics[width=3.3in]{rad1}\hspace{4mm}
%\caption{{\footnotesize The allowed range of the BD coupling parameter in $(r,\omega)$ parameter space
%for $n=r+1$ and small values of $r$ which is applicable for radiation dominated universe.}}
%\foreignlanguage{english}{\label{WE-rad1}}
%\end{figure}

 By using~(\ref{s-om}) and (\ref{d-gen-solution}) for this case, we obtain
\begin{eqnarray}\label{s-om-rad}
s&=&2(1-2r), \hspace{25mm} \omega=-\frac{5r-2}{2(2r-1)},\\\nonumber
\rho&=&\frac{3\phi_0(1-r)}{16\pi t_0^2}\left(\frac{t}{t_0}\right)^{-4r},\hspace{7mm}
 V=\frac{\phi_0 (1-r)(1-2r)}{t_0^2}\left(\frac{t}{t_0}\right)^{-4r}
\end{eqnarray}
For large values of the BD coupling parameter, we get $r=1/2$ and
then, $n=-1/2$ and $s=0$, namely
\begin{eqnarray}
a(t)\propto t^{1/2}, \hspace{10mm}
p(t)=3\rho(t)\propto t^{-2}\hspace{10mm} {\rm and} \hspace{10mm} \phi={\rm constant}.
\end{eqnarray}
Therefore, for the large values of $\mid\omega\mid$,
a radiation--dominated epoch evolves the same way as the
corresponding one of the flat FLRW space--time in GR.
Also the extra dimension contracts as $t$ increases.

We further see that
 for $0<r<1$, and $r\neq1/2$, the relations (\ref{ro-BD})
and (\ref{ro-phi}) indicate that for satisfying the weak energy
conditions, $\omega$ must be restricted
to $(4r-1)/(1-2r)<\omega<(-5r^2+6r-1)/(1-2r)^2$, in which for $r<1/2$, $\omega$
always takes positive values, while for $r>1/2$, it can be positive or negative.
Namely, for the case $n=r-1$, the model gives a decelerated universe consistent
with radiation dominated universe, in which the weak energy conditions are satisfied and
the fifth dimension contracts with the cosmic time.

Let us further describe the behavior of specific quantities for
different values of the cosmic time for the latter case. By assuming the obtained
ranges for $r$, $n$ and $s$, relations~(\ref{d-gen-solution})
imply that: i)  When $t$ goes to zero, by assuming $a_0=0=\phi_0$ at $t=0$, the scale factor $a$
%and the BD scalar field
goes to zero. However, the induced scalar potential,
the induced energy density and pressure, all take infinite
values. These behaviors indicate that the model is in accordance
with the big bang scenario. ii) As the cosmic time increases, the
scale factor
%and the scalar field
is increasing. However, the
expansion proceeds with a positive deceleration parameter. iii) When the
cosmic time tends to infinity, the scale factor
% as well as the BD scalar field
tends to infinity; while the other (induced) quantities,
such as the scalar potential and the energy density go to zero.
Therefore, the reduced cosmology in four--dimensional
space--time for $W_{_{\rm BD}}=1/3$ yields a model similar to the radiation
dominated obtained in GR.

\subsection{Vanishing Scalar Potential}
\label{Nariai}

As mentioned, in the quintessential scenarios or the early universe studies,
a few phenomenological models have been employed.
%the scalar potential in the BD setting can be added by hand to the action.
For example, by assuming two different models, the standard
BD theory has been generalized to include an effective cosmological constant:
%(let us consider the action as in footnote 6).
i) In some investigations, see, e.g.,~\cite{UK82}, an included scalar
potential can play the role of the cosmological constant.
This scalar potential can be reduced to a mass term or to a constant.
ii) Whilst in the other models~\cite{LS89,W89}, there is no scalar potential, but instead, there
is a perfect fluid with equation of state $P=-\rho$.
%For instance, in~\cite{RB92,RB93,Kol96}, such a
%constant or linear scalar potentials, together with (or without) a
%barotropic matter
%have been added to the BD action, and a few qualitative
%analyses have been presented.

As seen in the previous subsections, we studied the evolution of the
universe for some particular cases, having the general scalar
potential dictated from the geometry in $(D+1)$--dimensions.
However, from equation~(\ref{Eq-pot}), it is impossible to
 obtain a ${\rm constant}(\neq0)$ scalar potential, even by assuming a
 nonzero integration constant, for the
 power--law solution in the case where $s\neq2$.
 More precisely, by substituting the power--law
 solution~(\ref{power-law-phi}) into the corresponding scalar potential, the only way of having a
 constant scalar potential is to set $s=2$ which is in direct contradiction
 with the constraint $s\neq2$. (We should mention that having a
 scalar potential as a linear function of $\phi$ is also impossible.)
  In the other words, in our herein cosmological model, in which the scalar potential
 and the matter content are obtained geometrically from the extra dimensions,
imposing a few conditions on the parameters of the model,
 such that the induced scalar potential takes constant value (or a linear function of $\phi$)
 while the induced matter has barotropic equation of state,
  is in direct contradiction with having the power--law solutions for the spatially flat FLRW universe.
 We also should note that having an induced
 matter with $W_{\rm BD}=-1$, which can
 play the role of cosmological constant, is not allowed for $D=4$ dimensions.
 Although we have not investigated the case
 where $s=2$, we should also remark that, in this case, the logarithmic scalar
 potential cannot be a constant unless we set $t=constant$.
  Therefore, the only appropriate case is to consider the vanishing scalar
 potential which is obtained by setting $\omega=-1$ or $n=0$.
 In the case where $n=0$, we have $\psi={\rm constant}$, and thus, from Eqs.~(\ref{t-00}) and (\ref{t-ii}),
 we get $\rho_{_{\rm BD}}=0=p_{_{\rm BD}}$.
 It is straightforward to show that, in this case, we further
 obtain $r=1$, $s=D-2$, and $\omega=-(D-1)/(D-2)$.
 These values of parameters of the model lead us to
 an unknown zero acceleration universe.
 In the rest of this subsection we investigate the case of $\omega=-1$.

A particular case of cosmological power--law solutions is obtained
by assuming a vanishing scalar potential, $V(\phi)$. In our setting, this
type of solutions resembles the particular classes of Nariai solutions~\cite{N63,GFR73}
obtained for the four--dimensional flat FLRW
universe with $V(\phi)=0$ and
$\omega\neq4/[3(W^2-1)]<0$ (where $W$ is the equation of state parameter)
 for a perfect fluid
equation of state. By assuming $V(\phi)=0$, the
relations~(\ref{t-00}) and~(\ref{t-ii}) give
\begin{eqnarray}\label{t-00-zero.v}
p_{_{\rm BD}}=W_{_{\rm BD}}\rho_{_{\rm BD}}=-\frac{Crn}{t^{^{[1+(D-1)r+n]}}} \hspace{10mm} {\rm with}
\hspace{10mm} W_{_{\rm BD}}=\frac{r}{1-n} \hspace{10mm} {\rm and}
\hspace{10mm} C\equiv\frac{\phi_0}{8\pi}t_0^{^{[(D-1)r+n-1]}}.
\end{eqnarray}
On the other hand, equation~(\ref{v-def}) gives $\omega=-1$. Then,
from (\ref{s-om}), the exponent $n$ becomes
$n_{\pm}=\pm\sqrt{1-(D-1)r^2}$. In order to obtain real values for
$n$, we must have $\mid r \mid\leq1/\sqrt{D-1}$ for both values
of $n$. Furthermore, by assuming $C>0$, only $n_{-}$ guarantees
that the induced energy density takes positive values, and with
these conditions, the induced pressure is also positive.

As a particular case, let us discuss the radiative fluid, namely
$W_{_{\rm BD}}=1/(D-1)$. For this case, by applying $n_{-}$, we get $r=2/D$
and $n_{-}=-(D-2)/D$. By substituting these values of $r$ and $n_{-}$
in relation~(\ref{s-om}), the exponent of the cosmic time for the
BD scalar field reduces to $s=0$. In particular, when
$D=4$, we get the following solution
\begin{eqnarray}\label{Nar.rad}
a(t)\propto t^{1/2}, \hspace{10mm}
p_{_{\rm BD}}(t)=\frac{1}{3}\rho_{_{\rm BD}}(t)\propto
t^{-2}\hspace{10mm} {\rm and} \hspace{10mm} \phi={\rm constant},
\end{eqnarray}
which is similar to the Nariai solution derived for a radiative fluid in four dimensions.
As $n_{-}=-1/2$ (for $D=4$), thus the extra dimension contracts as the cosmic time increases.

Another case of particular interest is the solution
associated with cosmological constant, corresponding
to $W_{_{\rm BD}}=-1$. In BD gravity, such a solution is~not the
de~Sitter space, but it is the
power--law solution~\cite{MJ84,Faraoni.book}, which is a particular
case of the Nariai solution. The case of the Nariai
solution for $\omega>1/2$ is inflationary. This kind of inflation
is called power--law inflation in the context of GR, with an
exponential scalar field potential~\cite{AW84,LM85,Faraoni.book}.
By assuming a vanishing scalar potential, from
relation~(\ref{t-00-zero.v}) and again assuming $n_{-}$, we obtain
$r=-2/D$, $n=(D-2)/D$ and $s=2$. For a
four--dimensional space--time, we will have a decelerated expansion
and the BD coupling parameter is restricted to $\omega<1/2$.
\section{Conclusions}
\indent \label{conclusion}
 In this manuscript, we applied the dimensional reduction procedure,
%framework of the IMT toward
for a $(D+1)$--dimensional BD scenario.
Subsequently, we obtained a modified BD theory in $D$ dimensions.
Namely, with new dynamical features, more concretely, we have shown that, in this
scenario, the induced EMT, namely $T_{\mu\nu}^{^{[\rm BD]}}$ is
composed of three parts. The first part of the induced EMT, i.e.
$T_{\mu\nu}^{^{[\rm IMT]}}$, is the $(D+1)$th part of the metric,
which is geometrically induced on a hypersurface. Whereas, the second part,
$T_{\mu\nu}^{^{[\rm \phi]}}$, depends on the BD scalar field and
its derivatives with respect to the $(D+1)$th coordinate. The third
part is an induced scalar potential that can be derived from the
theory, and contributes in a wave
equation. In this construction, the induced EMT obeys a
conservation law. When the BD
scalar field takes constant values, the second and third parts of the induced EMT,
without loss of generality, can be set equal to zero and the
theory reduces to the GR setting derived in~\cite{RRT95} in
$D$--dimensions, as expected.
%Besides, when $\omega=-1$ the induced scalar potential vanishes.

Let us further emphasize some similarities and differences regarding a
standard $D$--dimensional BD theory: within the context of the
work at hand, the $(D+1)$--dimensional field equations
(\ref{(D+1)-equation-1}) and (\ref{(D+1)-equation-4}), with a
general metric (\ref{global-metric}), split naturally into four
sets of Eqs.~(\ref{D2say}), (\ref{BD-Eq-DD}), (\ref{D2-phi}) and
(\ref{P-Dynamic}), in which Eqs.~(\ref{BD-Eq-DD}) and
(\ref{D2-phi}) reproduce the BD field equations on a $D$--dimensional
space--time, with a geometrically induced energy--momentum source. Equivalently,
 they would be retrieved from a conventional action. Such a
correspondence is guaranteed by the Campbell--Magaard
theorem~\cite{C26,M63,RTZ95,LRTR97,SW03}. Whereas,
Eq.~(\ref{D2say}) has no BD analog, and the set of
Eqs.~(\ref{P-Dynamic}) is a generalized version of a conservation law introduced in the IMT.

We investigated solutions associated to the spatially flat FLRW in a vacuum
($D+1$)--dimensional space--time. By assuming a power--law
ansatz, (\ref{power-law-phi}), we found the general solutions for the equations.
Then, we discussed a few particular cases, such
as $(D +1)$--dimensional de Sitter--like space as well as an extended version of the O'Hanlon and Tupper
solution, which describes an empty universe in standard BD
theory.
%We also investigated the limits of these solutions when
%$D=4$, as $\omega$ goes to infinity and/or $\phi$ takes constant values.

We then employed the MBDT to study the ($D$--dimensional) induced setting. After
deriving the energy density, pressure and scalar
potential, we found that the general induced EMT describes a
perfect fluid. For power--law solutions, the induced
scalar potential is in the forms of the power--law or logarithmic.
However, we only considered the power--law case, which yielded
a baroscopic equation of state for the effective EMT.

We further calculated the $D$--dimensional energy density and pressure of the
BD scalar field, as well as density parameters associated to the
induced matter and BD scalar field.
Then, in order to meet recent observational data, as well as
to compare our results with the case from standard BD theory,
 we restricted the results to a four--dimensional space--time.
We should mention that the model has been constructed entirely from four parameters
$\omega$, $r$, $s$ and $n$, which are not independent. By imposing the different
physical conditions for each solution, we have obtained the allowed ranges of the present parameters
of the model.

In order to discuss on the extended quintessence of the dark energy
models and consequently having an accelerating universe,
 the parameter $r$ must be taken greater than one.
By assuming this condition for $r$, we found that $s$ is
always negative whereas $n$ can take positive as well as
negative values. However, as the negative values of this parameter give
a contracting fifth dimension, we restricted the solutions by omitting
the solutions associated to positive values of $n$. Further,
we also restricted the ranges of the parameters so that weak energy conditions for the
induced matter, as well as matter associated to BD scalar field, would be satisfied.
These physical conditions give a permissable range for the BD coupling parameter presented by
(\ref{omega-s.quint-2}) or equivalently by (\ref{omega-n.quint-2}).
By applying the physical conditions, for an accelerating expansion,
the allowed ranges of $\omega$ have been plotted versus $r$ and $n$.
The resulted ranges for $\omega$ were further constrained to match with other limits,
reported by recent observational data for the parameters of the model.
Finally, as Fig.~\ref{quint-obser-omega-r-n.fig} shows, we found a narrow region of
the BD coupling parameter in the $(r,\omega)$ parameter space, for small negative values
of $n$ and permitted ranges of the density parameters obtained from observations.
However, these values for $\omega$
are~not in accordance with the constraints on $\omega$ reported
by the tests of the solar system\rlap.\footnote{Very stringent constraints
have been put on the BD model
by means of a few solar system experiments~\cite{W06}.
For example, the constraint on the BD coupling
parameter indicated that it should be restricted to $\omega>40000$~\cite{BIT03}.}
Although the allowed ranges of the BD
coupling parameter in our model are very similar to those obtained in
investigations in the context of conventional BD theory, we should stress that
our herein model leads us to accelerating solutions based on fundamental concept, i.e.,
from the geometrical presence of the extra dimension, rather than by means of some {\it ad hoc}
assumptions.

We also probed the variation of the gravitational coupling and the age of the universe.
The result is in accordance with the observational data~\cite{LWC13}.
In our model, as $s$ takes negative values,
the gravitational coupling increases with cosmic time.

In order to proceed the discussions on the dust and
radiation dominated universe, we further
obtained the decelerated parameter, $q$, and expressed all the
induced quantities in terms of the parameter of the equation of
state associated to the induced matter, $W_{_{\rm BD}}$,
the exponent associated to the scale factor, $r$, the
number of the dimensions, $D$, and the cosmic time~$t$
for a few well--known particular cases of the equation of
state, namely, matter--dominated and radiation--dominated
universe. The results of the model have shown that the
cosmological model based on the MBDT
can describe, by means of assuming a few required
physical conditions, an accelerated universe as well as
a decelerated universe.
We have plotted
%, according to required physical conditions,
the allowed ranges of BD coupling parameter.
Also, in Fig.~\ref{wphi-r-matt}, for a
particular value of $\omega$ associated to the matter--dominated case,
we have specified $W_{\phi}$ for small values of $r$, which shows a proper
range consistent with the recent observational results.

For a dust fluid, we also argued that the terms inside the brackets of
equation (\ref{DD-FRW-shetab}), which represents a
quantity associated to the strong energy condition, decides to have
a decelerating or an accelerating expansion.
For the particular case of the dust in four dimensions, by expressing
all the quantities versus $n$ and $s$, we found that when a nonzero $s$ is
restricted to the ranges $-1<s<2$ and $s<-1$, then the universe is
in the phase of deceleration and acceleration, respectively.
For $s<-1$, the deceleration parameter is always restricted to the range $-1<q<0$, while for
$-1<s<2$, it takes positive values.
In both of these cases, the total energy density satisfies the weak energy condition.
%In the mentioned cases, a varying induced scalar
%potential, corresponding to each epoch, contributes to describe dark matter and dark energy.
%This identity of the MBDT is completely different with the phenomenological models, based on BD
% theory, in which the scalar potential (which is introduced by hand) must be a constant value (or zero).
 Also, a zero
acceleration phase is described when $s=-1$ (corresponds to $r=1$), in which also the
weak energy condition is satisfied.
Consequently, we found that in the case where $n=-1$, the herein
cosmological model can be considered as an
unified model for dark matter--dark energy.
% in which the
 %induced scalar potential varying with time and assists
 %to construct a proper model corresponding to each epoch.
 We also mention that, in this case, the extra
 dimension contracts as the cosmological time increases.
 We should also note that, similar to the most of
 the conventional models of BD theory, the BD coupling parameter takes small negative
values, which is, as mentioned, in contradiction with solar system experiments.
Different phenomenological approaches have been trying to resolve the problems with the
cosmological models in the BD theory, especially the trouble with $\omega$~\cite{LSB89,BM90,PSW08,WW12}.
One way to get out of this problem can be
assuming a varying $\omega$, namely, $\omega=\omega(\phi)$
instead of the constant one. In the context of MBDT, unfortunately,
using such an approach would flunk~\cite{Ponce1}.
We should also stress that the observational
constrains on $\omega$ are only in the weak field limit.

It would be of interest to study a case in which a scalar
potential can play the role of an effective
cosmological constant.
However, as in the MBDT, both the induced matter and the scalar potential
are dictated from the geometry of the extra dimension, thus, in the power--law cosmological
setting associated to the MBDT, neither such a scalar
potential nor an induced matter with $W_{\rm BD}=-1$
is obtained.
%a fully observationally viable model, by
%assuming several reasonable values for the parameters present.
%could be interesting to proceed the
%calculations for the case where the induced scalar
%potential $V(\phi)$ (with a positive constant value) can play the role of cosmological constant.
%However, as the induced matter and scalar potential
%dictated from the geometry of the extra dimension rather than by hand, the cosmological setting
%based on the MBDT does not allow us
 %Instead, as mentioned, the induced scalar potential in MBDT
%varies by time and automatically leads to different phases of the universe.
Instead, we can investigate the case where the scalar potential is zero and obtain the
corresponding cosmological phases.
By assuming a vanishing induced scalar potential, we
studied a particular case of cosmological power--law solutions
resembling to the Nariai class~\cite{N63,GFR73}, obtained for the
four--dimensional flat FLRW universe with $V(\phi)=0$ and
$\omega\neq4/[3(W^2-1)]<0$, for a perfect fluid equation of state.
We also found
that, in the other special case where $n=0$, the scalar potential
vanishes and leads to an unknown zero acceleration universe.

\section*{Acknowledgments}
We would like to thank an anonymous referee for the fruitful comments.
SMMR appreciates for the support of grant SFRH/BPD/82479/2011 by
the Portuguese Agency Funda\c{c}\~ao para a Ci\^encia e
Tecnologia. This research
is supported by the grants CERN/FP/123618/2011 and PEst-OE/MAT/UI0212/2014.

\appendix
\section{$(D+1)$-dimensional BD theory and the Kaluza-Klein theory}
\label{App.A}

As mentioned at the end of section~\ref{Set up},
we can derive the BD theory by applying
%the Lyra geometry or
the classical KK theory.
Here, we derive a $(D+1)$--dimensional BD theory from a KK theory by extending the
 usual approach applied to obtain the $4$--dimensional BD
theory~\cite{OW97, scalarbook, Faraoni.book, PS02}.

We consider a KK theory with a single
dilaton field. Let us be more clear. We start from a $(D+1+d)$--dimensional
space-time\footnote{For denoting the $(D+1+d)$ quantities,
we will use a caret.} $(M\otimes N, \hat{\Upsilon})$,
where $M$ and $N$ are a $(D+1)$--dimensional manifold
(with one timelike dimension) and a submanifold with
$d$ ($d\geq1)$ spatial dimensions, respectively.
We assume that the $(D+1+d)$--dimensional metric
with $d$--dimensional compactified space is
in the form
\begin{eqnarray}\label{KK-metric}
(\hat{\Upsilon}_{AB})=\left(
  \begin{array}{cc}
    \hat{\gamma}_{ab}(x) & 0 \\
    0 & \hat{\phi}_{\bar{a}\bar{b}} \\
  \end{array}
\right),
\end{eqnarray}
where $A, B, ...=0,1,2,..., (D+d)$, $a,b,..=0,1,2,...,D$,
$\bar{a},\bar{b},...=(D+1), (D+2),...,(D+d)$
 and $x$ denotes the coordinates of $(D+1)$--dimensional space--time.

In the original KK theory~\cite{{OW97}}, three key assumptions have been considered within a metric
similar to~(\ref{KK-metric}), but a significant difference, namely, with non--vanishing off--diagonal
terms $A_a$ corresponding to the electromagnetic potentials. However, here, we also assume
the three key assumptions of the original KK theory, but instead,
%the same as the KK models in cosmology,
we assume that
there are no off-diagonal terms in the above metric.
Furthermore, in order to respect to homogeneity and
isotropy, we assume that $\hat{\gamma}_{ab}$ is a generalized FLRW
metric on the $(D+1)$--dimensional manifold
$M$ and $\hat{\phi}_{\bar{a}\bar{b}}$ is a diagonal Riemannian metric on $N$. Moreover, we assume that
$\hat{\phi}_{\bar{a}\bar{b}}=\ell^2(x)\tilde{\phi}_{\bar{a}\bar{b}}(\theta)$
 (where we have chosen $\theta$'s as dimensionless coordinates)
where the purely geometrical metric $\tilde{\phi}_{\bar{a}\bar{b}}$ describes
the $d$-dimensional internal space
and $\ell(x)$, which is a function of $(D+1)$--dimensional coordinate $x$,
stands for the size of the space.
Thus, we have $\sqrt{-\hat{\Upsilon}}=\sqrt{-\hat{\gamma}}A^d\sqrt{\tilde{\phi}}$ in which
$\sqrt{\tilde{\phi}}$ is related to the volume of the compact manifold $N$, $V_d=\ell^d\tilde{V}_d$, via
$\tilde{V}_d=\int\sqrt{\tilde{\phi}}d^d\theta$, where $\hat{\Upsilon}$, $\hat{\gamma}$
and $\tilde{\phi}$ are the determinant of the corresponding metrics.

By adapting the KK first key assumption, the action associated to a $(D+1+d)$--dimensional
space-time is written as
\begin{equation}\label{KK-action-1}
S=\frac{1}{16\pi \hat{G}}\int d^{^{(D+1+d)}}x \sqrt{-\hat{\Upsilon}}\hat{R},
\end{equation}
where $\hat{R}$ stands for the Ricci curvature
associated to the metric $\hat{\Upsilon}_{AB}$ and $\hat{G}$ denotes the
$(D+1+d)$--dimensional gravitational coupling.
In order to derive the effective action on $M$, we can compute the integral over the
$D+1+d$ dimensions by assuming that the above integral is
composed of multiplication of two integrals, one over
the $(D+1)$ dimensions and the other over $d$ dimensions.
Besides, by defining \cite{Faraoni.book} $\varphi\equiv\mid\!\!\det(\hat{\phi}_{\bar{a}\bar{b}})\!\!\mid$
and a symmetric tensor $\varrho_{\bar{a}\bar{b}}\equiv\varphi^{-1/d}\phi_{\bar{a}\bar{b}}$
such that $\mid\det(\varrho_{\bar{a}\bar{b}})\mid=1$,
we get
\begin{equation}\label{KK-action-2}
S=\frac{1}{16\pi G}\int d^{^{(D+1)}}x \sqrt{-{\gamma}}\left[\phi\left(R+R_{_{\rm N}}\right)
+\frac{(d-1)}{d}\frac{{\gamma}^{ab}\,(\nabla_a\phi)(\nabla_b\phi)}{\phi}\right],
\end{equation}
where $R_{_{\rm N}}$ is the Ricci curvature associated
to the $d$--dimensional submanifold $N$ and we have set
$\phi=\sqrt{\varphi}$ and $G=\hat{G}/V_d$.
By comparing~(\ref{(D+1)-action}) and (\ref{KK-action-2}), we find that the
$(D+1)$--dimensional BD action for the vacuum case can be
derived from a $(D+d+1)$ extended GR (or the simplest KK theory) by
setting the BD coupling parameter as $G=1$, $\omega=-1+1/d$ in which $d$ is the number of
compact extra spatial dimensions and the BD scalar
field emerges geometrically from the determinant of the
submanifold $N$.

%%%%%%%%%%%%%%%%%%%%%%%%%%%%%%%%%%%%%%%%%%%%%%%%%%%%%%%%%%%%%%%%%%%%%%%%%%%%%%%%%%%%%%%%%%%%%%%%%%%%%%%%%%%

\end{document}